\documentclass{nature}

\usepackage{times}
\usepackage{xcolor, amssymb, gensymb, graphicx, siunitx, geometry}
\usepackage{url}
\usepackage{comment}

\makeatletter
\let\saved@includegraphics\includegraphics
\AtBeginDocument{\let\includegraphics\saved@includegraphics}
\renewenvironment*{figure}{\@float{figure}}{\end@float}
\makeatother

\title{Rapid identification of pathogenic bacteria using Raman spectroscopy and deep learning}

\author
{Chi-Sing Ho,$^{1,2,\ast,\Diamond}$
Neal Jean,$^{3,4,\ast}$ 
Catherine A. Hogan,$^{5,6}$ 
Lena Blackmon,$^{2}$
Stefanie S. Jeffrey,$^{7}$\\
Mark Holodniy,$^{8,9,10}$ 
Niaz Banaei,$^{5,6,10}$
Amr A. E. Saleh,$^{2,11,\Diamond}$ 
Stefano Ermon,$^{3,\Diamond}$ and 
Jennifer Dionne$^{2,\Diamond}$\\
\\
\normalsize{$^{1}$Dept. of Applied Physics, Stanford University, Stanford, CA}\\
\normalsize{$^{2}$Dept. of Materials Science and Engineering, Stanford University, Stanford, CA}\\
\normalsize{$^{3}$Dept. of Computer Science, Stanford University, Stanford, CA}\\
\normalsize{$^{4}$Dept. of Electrical Engineering, Stanford University, Stanford, CA}\\
\normalsize{$^{5}$Dept. of Pathology, Stanford University School of Medicine, Stanford, CA}\\
\normalsize{$^{6}$Clinical Microbiology Laboratory, Stanford Health Care, Stanford, CA}\\
\normalsize{$^{7}$Dept. of Surgery, Stanford University School of Medicine, Stanford, CA}\\
\normalsize{$^{8}$Dept. of Medicine, Stanford University School of Medicine, Stanford, CA}\\
\normalsize{$^{9}$}VA Palo Alto Health Care System, Palo Alto, CA\\
\normalsize{$^{10}$Division of Infectious Diseases and Geographic Medicine, Dept. of Medicine, Stanford University School of Medicine, Stanford, CA}\\
\normalsize{$^{11}$Dept. of Engineering Mathematics and Physics, Faculty of Engineering, Cairo University, Giza, Egypt}\\
\normalsize{$^{\ast}$These authors contributed equally to this manuscript}\\
\\
\normalsize{$^\Diamond$To whom correspondence should be addressed; E-mail:  csho@alumni.stanford.edu, aessawi@stanford.edu, ermon@cs.stanford.edu, jdionne@stanford.edu.}
\\
}

\definecolor{darkgreen}{rgb}{0.01, 0.75, 0.24}

\date{}

\begin{document} 


\baselineskip24pt


\maketitle



\newpage
\clearpage
\pagebreak

\begin{abstract}
Raman optical spectroscopy promises label-free bacterial detection, identification, and antibiotic susceptibility testing in a single step. However, achieving clinically relevant speeds and accuracies remains challenging due to weak Raman signal from bacterial cells and numerous bacterial species and phenotypes. Here we generate an extensive dataset of bacterial Raman spectra and apply deep learning approaches to accurately identify 30 common bacterial pathogens. Even on low signal-to-noise spectra, we achieve average isolate-level accuracies exceeding 82\% and antibiotic treatment identification accuracies of 97.0$\pm$0.3\%. We also show that this approach distinguishes between methicillin-resistant and -susceptible isolates of \textit{Staphylococcus aureus} (MRSA and MSSA) with 89$\pm$0.1\% accuracy. We validate our results on clinical isolates from 50 patients. Using just 10 bacterial spectra from each patient isolate, we achieve treatment identification accuracies of 99.7\%. Our approach has potential for culture-free pathogen identification and antibiotic susceptibility testing, and could be readily extended for diagnostics on blood, urine, and sputum.

\end{abstract}


\newpage
\clearpage
\pagebreak


\section*{Introduction}
Bacterial infections are a leading cause of death in both developed and developing nations, taking more than 6.7 million lives each year\cite{Fleischmann2016-uc, DeAntonio2016-en}.
These infections are also costly to treat, accounting for 8.7\% of annual healthcare spending, or \$33 billion, in the United States alone\cite{Torio2016-vz}.
Current diagnostic methods require sample culturing to detect and identify the bacteria and its antibiotic susceptibility, a slow process that can take days even in state-of-the-art labs\cite{Dellinger2013-xq,Chaudhuri2008-ac}.
Broad spectrum antibiotics are often prescribed while waiting for culture results\cite{American_Thoracic_Society2005-xy}, and according to the Centers for Disease Control and Prevention, over 30\% of patients are treated unnecessarily\cite{Fleming-Dutra2016-oq}.
New methods for rapid, culture-free diagnosis of bacterial infections are needed to enable earlier prescription of targeted antibiotics and help mitigate antimicrobial resistance.

Raman spectroscopy has the potential to identify the species and antibiotic resistance of bacteria, and when combined with confocal spectroscopy, can interrogate individual bacterial cells (Figure 1a, b).
Different bacterial phenotypes are characterized by unique molecular compositions, leading to subtle differences in their corresponding Raman spectra. 
However, because Raman scattering efficiency is low ($\sim10^{-8}$ scattering probability\cite{Butler2016-lb}), these subtle spectral differences are easily masked by background noise. 
High signal-to-noise ratios (SNRs) are thus needed to reach high identification accuracies\cite{Stockel2016-bg}, typically requiring long measurement times that prohibit high-throughput single-cell techniques.
Additionally, the large number of clinically relevant species, strains, and antibiotic resistance patterns require comprehensive datasets that are not gathered in studies that focus on differentiating between species\cite{Kloss2013-vr,Boardman2016-tp}, isolates (typically referred to as strains in the literature)\cite{Schmid2009-xt,Munchberg2014-ee}, or antibiotic susceptibilities\cite{Novelli-Rousseau2018-zo,Liu2016-vo,Lu2013-sd,Germond2018-ju,Ayala2018-ie,Kirchhoff2018-ho}. In this work, we address this challenge by training a convolutional neural network (CNN) to classify noisy bacterial spectra by isolate, empiric treatment, and antibiotic resistance.


\section*{Results}

\section*{Deep learning for bacterial classification from Raman spectra}
In order to gather a training dataset, we measure Raman spectra using short measurement times on dried monolayer samples, as illustrated in Figure 1. We ensure that the majority of individual spectra are taken over single cells and preparation conditions are consistent between samples (See Methods). We construct reference datasets of 60,000 spectra from 30 bacterial and yeast isolates for 3 measurement times --- these 30 isolate classes cover over 94\% of all bacterial infections treated at Stanford Hospital in the years 2016-17 and are representative of the majority of infections in intensive care units worldwide\cite{Vincent2009-kq}. We further augment our reference dataset with 12,000 spectra from clinical patient isolates, including MRSA and MSSA isolates (see Methods for full dataset information). Previously, the lack of large datasets prohibited the use of CNNs due to the high number of spectra per bacterial class needed for training.

In recent years, CNNs have been applied with tremendous success to a broad range of computer vision problems\cite{Krizhevsky2012-ke,Mnih2014-ei,Karpathy2015-it,Zhang2016-yq,Dong2014-iy,Wang2015-va,Girshick2014-cv,Kraus2018-su,Lotfollahi2019-bf,Berisha2019-gr}.
However, while classical machine learning techniques have been applied to spectral data\cite{Boardman2016-tp,Schmid2009-xt,Novelli-Rousseau2018-zo,Kampe2017-ue,Guo2018-iw}, relatively little work has been done in adapting deep learning models to spectral data\cite{Gurbani2018-da,Malek2018-cp,Liu2017-wk,Zhang2019-ju}. In particular, state-of-the-art CNN techniques from image classification such as residual connections have previously not been applied to low SNR, 1D spectral data.
Our CNN architecture consists of 25 1D convolutional layers and residual connections\cite{He2015-zl} --- instead of two-dimensional images, it takes one-dimensional spectra as input (see Methods for further detail).
Unlike previous work, we do not use pooling layers and instead use strided convolutions with the goal of preserving the exact locations of spectral peaks\cite{Dumoulin2016-bx}.
Empirically, we find that this strategy improves model performance.

We train the neural network on a 30-class isolate identification task, where the CNN outputs a probability distribution across the 30 reference isolates and the maximum is taken as the predicted class.
The model is trained on the reference dataset and tested on an independent test dataset gathered from separately cultured samples.

A performance breakdown for individual classes is displayed in the confusion matrix in Figure 2a. Here, we show data for 1 s measurement times, corresponding to a SNR of 4.1 --- roughly an order of magnitude lower than typical reported bacterial spectra\cite{Boardman2016-tp,Schmid2009-xt,Kloss2013-vr}; classification accuracies increase with SNR, as shown in Supplementary Figure 1. On the 30-class task, the average isolate-level accuracy is $82.2\pm0.3$\% ($\pm$ calculated as standard deviation across 5 train and validation splits). Gram-negative bacteria are primarily misclassified as other Gram-negative bacteria; the same is generally true for Gram-positive bacteria, where additionally, the majority of misclassifications occur within the same genus. 
In comparison, our implementations of the more common classification techniques of logistic regression and support vector machine (SVM) achieve accuracies of 75.7\% and 74.9\%, respectively.

\section*{Identification of empiric treatments and antibiotic resistance}
Species-level classification accuracy is the standard metric for bacterial identification, but in practice, the priority for physicians is choosing the correct antibiotic to treat a patient.
Common antibiotics often have activity against multiple species, so the 30 isolates can be arranged into groupings based on the recommended empiric treatment if the bacterial species is known.
Classification accuracies can thus be condensed into a new confusion matrix grouped by empiric antibiotic treatment (Figure 2b), where the average accuracy of our method is 97.0$\pm$0.3\%. In comparison, logistic regression and SVM achieve accuracies of 93.3\% and 92.2\%, respectively.

Beyond empiric first choice antibiotics, clinicians also conduct antibiotic susceptibility tests to determine bacterial responses to drugs. As a step toward a culture-free antibiotic susceptibility test using Raman spectroscopy, we  train a binary CNN classifier to differentiate between methicillin-resistant and -susceptible isolates of \textit{S. aureus}. This model achieves 89.1$\pm$0.1\% identification accuracy (Figure 3a). Because the consequences for misdiagnosing MRSA as MSSA are often more severe than the reverse misdiagnosis, the binary decision can be tuned for higher sensitivity (low false negative rate), as shown in the receiver operating characteristic (ROC) curve in Figure 3b (dotted line denotes performance of random guessing). The area under the curve (AUC) is 0.953, meaning that a randomly selected positive example (i.e., Raman sample from patient with MRSA) will be predicted to be more likely to be MRSA than a randomly selected negative example (i.e., sample from patient with MSSA) with probability 0.953.

\section*{Extension to clinical patient isolates}
To demonstrate that this approach can be extended to new clinical settings, we test our model on two groups of 25 clinical isolates derived from patient samples, for a total of 50 patients, Within each patient group, samples include 5 isolates from each of the 5 most prevalent\cite{Banaei_undated-ux} empiric treatment groups (see Supplementary Table 2 and Supplementary Figure 4). We first consider isolates from 25 patients collected from Palo Alto VA Medical Center in 2018. We augment our reference dataset with this clinical dataset comprised of 400 spectra per clinical isolate.
To account for changes in the relative prevalence of species and antibiotic resistances over time, the model may be fine-tuned on a small dataset that is representative of current patient populations.
We use a leave-one-patient-out cross-validation (LOOCV) strategy for fine-tuning, where we assign 1 patient in each class to the test set (5 patients total) and use the other 4 for fine-tuning (20 patients total), fine-tuning on 10 randomly sampled spectra per patient isolate --- we repeat this process 5 times, so all 25 patient isolates appear in the held-out test set once.
We then use 10 randomly sampled spectra from each patient isolate in the test set to reach an infection identification for that patient isolate.
The sampling procedure for identification is repeated for 10,000 trials, and we report the average accuracy and standard deviation, and display a trial representing the modal result in Figure 4a (full experiment details can be seen in Supplementary Note 1). 
A CNN pre-trained on the reference dataset serves both as initialization for the fine-tuned model and as a baseline, achieving 89.0$\pm$3.6\% ($\pm$ calculated as standard deviation across 10,000 sampling trials) species identification accuracy, a statistically significant improvement over logistic regression and support vector machine baselines (see Methods for details).
When the CNN is fine-tuned on clinical data and then evaluated on the held-out patients, the identification accuracy is improved to 99.0$\pm$1.9\% (Supplementary Figure 5).
Samples for the clinical tests were prepared separately for each patient, so we conclude that the measured performance is not due to batch effects from sample preparation or measurement conditions.

Because patient samples may contain very low numbers of bacterial cells without culturing (e.g. 1 CFU/mL or fewer in blood\cite{Lamy2016-vs}), only a few individual bacterial spectra per patient may be available to make a diagnosis. As seen in Figure 4c, just 10 cellular spectra are enough to reach high identification accuracy.
The rate of correct identification using 10 spectra is 99.0\%, within 1\% of the performance with 400 spectra (100.0\%).
While acquiring spectra from 400 individual bacterial cells would likely necessitate culturing, we achieve high accuracy on spectra from 10 individual bacterial cells, commensurate with typical levels of bacterial cells present in uncultured samples\cite{Lamy2016-vs,Reimer1997-rr}.

For a proof-of-concept antibiotic susceptibility test on clinical isolates, we collect Raman spectra on 5 additional clinical MRSA isolates and test the binary MRSA/MSSA classifier that is pre-trained on the reference MRSA and MSSA isolates. Using the same LOOCV process, we fine-tune the binary classifier on the clinical spectra. A representative result is shown in Figure 4b; any misclassifications of MSSA as MRSA are labeled as ``suboptimal'', indicating that Vancomycin (prescribed for MRSA) is also effective on MSSA but is not considered optimal treatment and may introduce adverse patient effects. On average, the pre-trained binary classifier achieves 61.7$\pm$7.3\% accuracy and the fine-tuned binary classifier achieves 65.4$\pm$6.3\% accuracy (Supplementary Figure 5).

Finally, to test the robustness of the fine-tuning approach over multiple clinical datasets, we use our second patient group of 25 isolates, collected from Stanford Hospital from February 2019 to March 2019. We conduct additional fine-tuning of the model that is pre-trained on the reference dataset and fine-tuned on the original clinical dataset. The treatment group identification accuracy on the new clinical dataset using only 10 spectra per patient is 99.7$\pm$1.1\% Figure 4 d, e, with improved performance for both \textit{S. aureus} and \textit{P. aeruginosa}, demonstrating the potential for continuous improvement of the trained model.


\section*{Discussion}
In this work, we apply state-of-the-art deep learning techniques to noisy Raman spectra to identify clinically relevant bacteria and their empiric treatment.
A CNN model pre-trained on our dataset can easily be extended to new clinical settings through fine-tuning on a small number of clinical isolates, as we have shown on our clinical dataset. We envision that fine-tuning processes such as the one demonstrated here could be important components for continuously evaluating and improving deployed models.
Our model, applied here to the identification of clinically relevant bacteria, can be applied with minimal modification to other identification problems such as materials identification, or other spectroscopic techniques such as nuclear magnetic resonance, infrared, or mass spectrometry.

This study uses measurement times of 1 s, corresponding to SNRs that are an order of magnitude lower than typical reported bacterial spectra --- while still achieving comparable or improved identification accuracy on more isolate classes than typical Raman bacterial identification studies.
A common strategy for reducing measurement times is surface-enhanced Raman scattering (SERS) using plasmonic structures, which can increase the signal strength by several orders of magnitude\cite{Boardman2016-tp,Kogler2018-pf,Chen2018-zr}. SERS spectra can be highly variable and difficult to reproduce, particularly on cell samples\cite{Li2010-oc,Butler2016-lb}, making it difficult to develop a reliable diagnostic method based on SERS.
However, with a dataset capturing the breadth of variation in SERS spectra, a CNN could enable a platform that processes blood, sputum, or urine samples in a few hours.

Compared to other culture-free methods\cite{Cronquist2012-uu} including single-cell sequencing\cite{Kang2014-vy,Tung2017-uo,Wang2015-fd,Pallen2010-jo} and fluorescence or magnetic tagging\cite{Chung2015-zk}, Raman spectroscopy has the unique potential to be a technique for identifying phenotypes that does not require specially designed labels, allowing for easy generalizability to new strains.

To achieve treatment recommendations as fine-grained as those from culture-based methods, larger datasets covering more resistant and susceptible clinical isolates, greater diversity in antibiotic susceptibility profiles, cell states, and growth media and conditions would be needed.
Though collecting such datasets is beyond an academic scope, requiring highly automated sample preparation and data acquisition processes, there is promise for clinical translation.
Similarly, studies applying the Raman-CNN system to identify pathogens in relevant biofluids such as whole blood, sputum, and urine are a promising future direction to demonstrate the validity of the method as a diagnostic tool.
When combined with such an automated system, the Raman-CNN platform presented here could rapidly scan and identify every cell in a patient sample and recommended an antibiotic treatment in one step, without needing to wait for a culture step. Such a technique would allow for accurate and targeted treatment of bacterial infections within hours, reducing healthcare costs and antibiotics misuse, limiting antimicrobial resistance, and improving patient outcomes.

\newpage
\clearpage


\newpage
\clearpage
\begin{figure} \centering \spacing{1}
\includegraphics[width=1\textwidth]{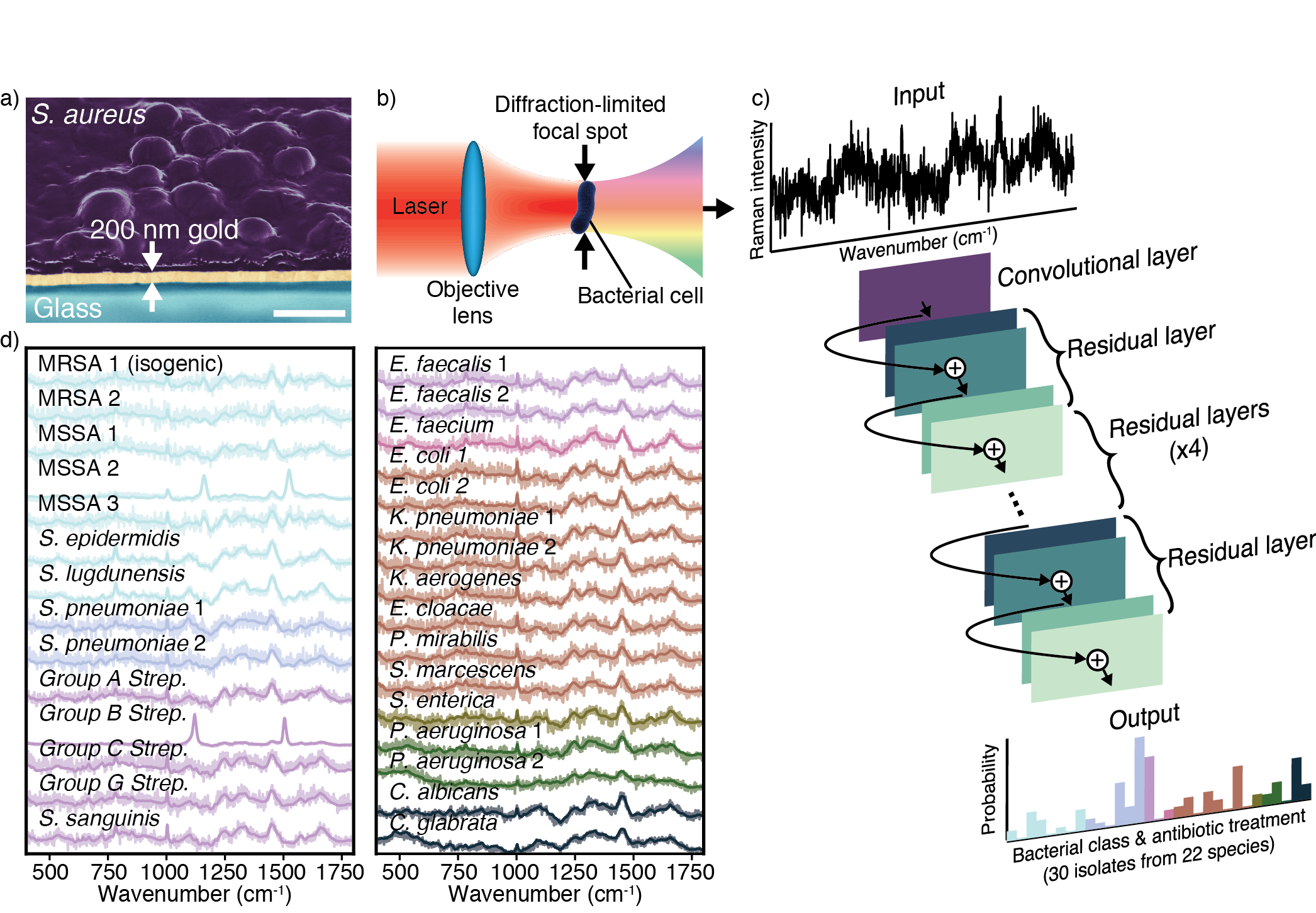}
\caption{A convolutional neural network (CNN) can be used to identify bacteria from Raman spectra.
a) To build a training dataset of Raman spectra, we deposit bacterial cells onto gold-coated silica substrates and collect spectra from 2000 bacteria over monolayer regions for each strain. An SEM cross section of the sample is shown (gold coated to allow for visualization of bacteria under electron beam illumination). Scale bar is 1 $\mu$m.
b) Conceptual measurement schematic: by focusing the excitation laser source to a diffraction-limited spot size, Raman signal from single cells can be acquired.
c) Using a one-dimensional residual network with 25 total convolutional layers (see Methods for details), low-signal Raman spectra are classified as one of 30 isolates, which are then grouped by empiric antibiotic treatment.
d) Raman spectra of bacterial species can be difficult to distinguish, and short integration times (1 s) lead to noisy spectra (SNR = 4.1). Averages of 2000 spectra from 30 isolates are shown in bold and overlaid on representative examples of noisy single spectra for each isolate. Spectra are color-grouped according to antibiotic treatment. These reference isolates represent over 94\% of the most common infections seen at Stanford Hospital in the years 2016-17\cite{Banaei_undated-ux}.
}
\label{fig:diagram}
\end{figure}

\newpage
\clearpage
\begin{figure} \centering \spacing{1}
\includegraphics[width=1\textwidth]{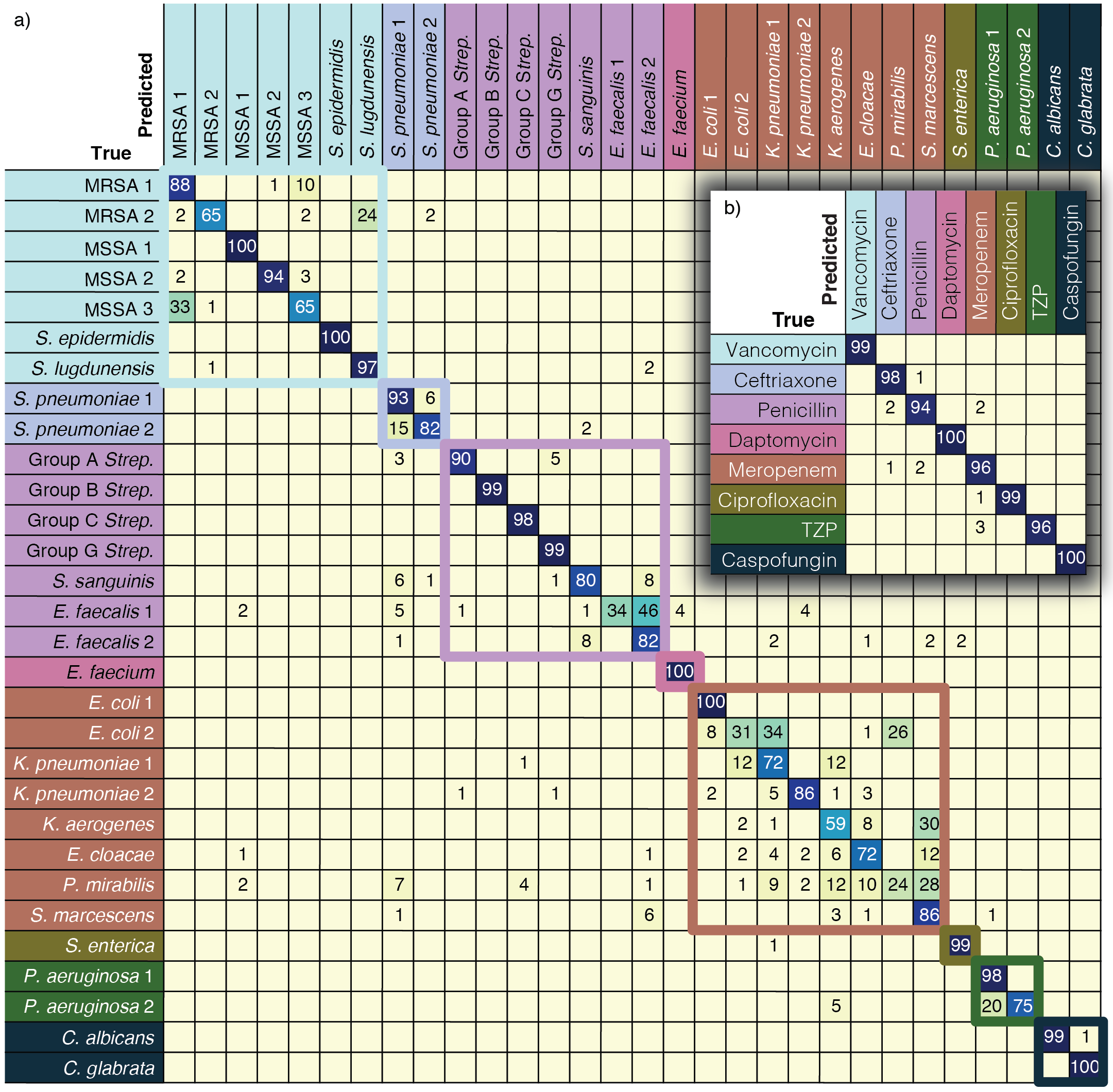}
\caption{CNN performance breakdown by class. The trained CNN classifies 30 bacterial and yeast isolates with isolate-level accuracy of 82.2$\pm$0.3\% and antibiotic grouping-level accuracy of 97.0$\pm$0.3\% ($\pm$ calculated as standard deviation across 5 train and validation splits).
a) Confusion matrix for 30 strain classes. Entry (i, j) represents the percentage out of 100 test spectra that are predicted by the CNN as class j given a ground truth of class i; entries along the diagonal represent the accuracies for each class. Misclassifications are mostly within antibiotic groupings, indicated by colored boxes, and thus do not affect the treatment outcome. Values below 0.5\% are not shown, and matrix entries covered by figure insets are all below 0.5\% aside from a 2\% misclassification of MRSA 2 as \textit{P. aeruginosa} 1 and 1\% misclassification of Group B \textit{Strep}. as \textit{K. aerogenes}.
b) Predictions can be combined into antibiotic groupings to estimate treatment accuracy. TZP = piperacillin-tazobactam. All values below 0.5\% are not shown.}
\label{fig:cm}
\end{figure}

\newpage
\clearpage
\begin{figure} \centering \spacing{1}
\includegraphics[width=0.5\textwidth]{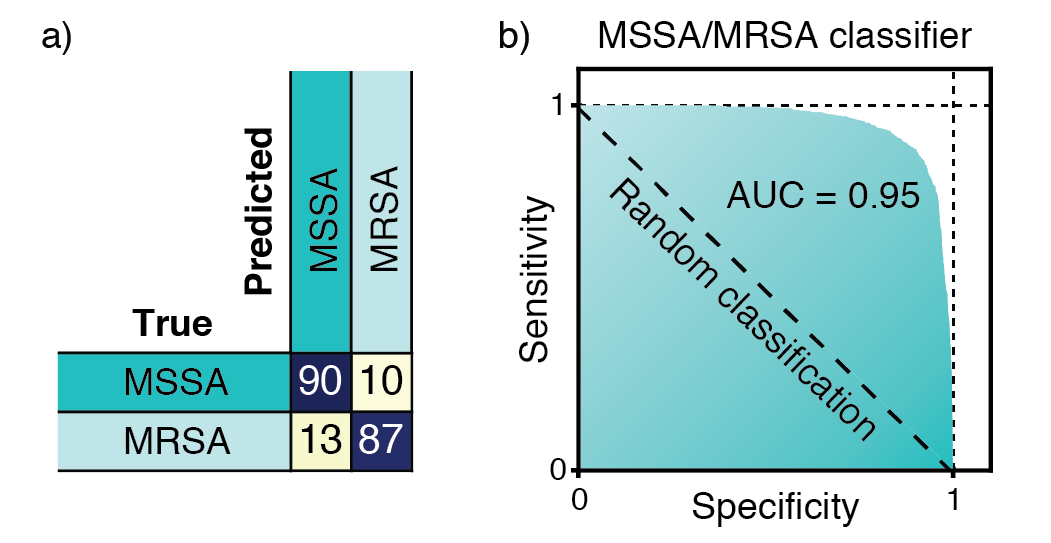}
\caption{Binary MRSA/MSSA classifier.
a) A binary classifier is used to distinguish between methicillin-resistant and -susceptible \textit{S. aureus} (MRSA/MSSA), achieving 89.1$\pm$0.1\% accuracy.
b) By varying the classification threshold, it is possible to trade off between sensitivity (true positive rate) and specificity (true negative rate). The ROC curve shows sensitivities and specificities significantly higher than random classification, with an AUC of 0.953.
}
\label{fig:mssamrsa}
\end{figure}

\newpage
\clearpage
\begin{figure} \centering \spacing{1}
\includegraphics[width=1\textwidth]{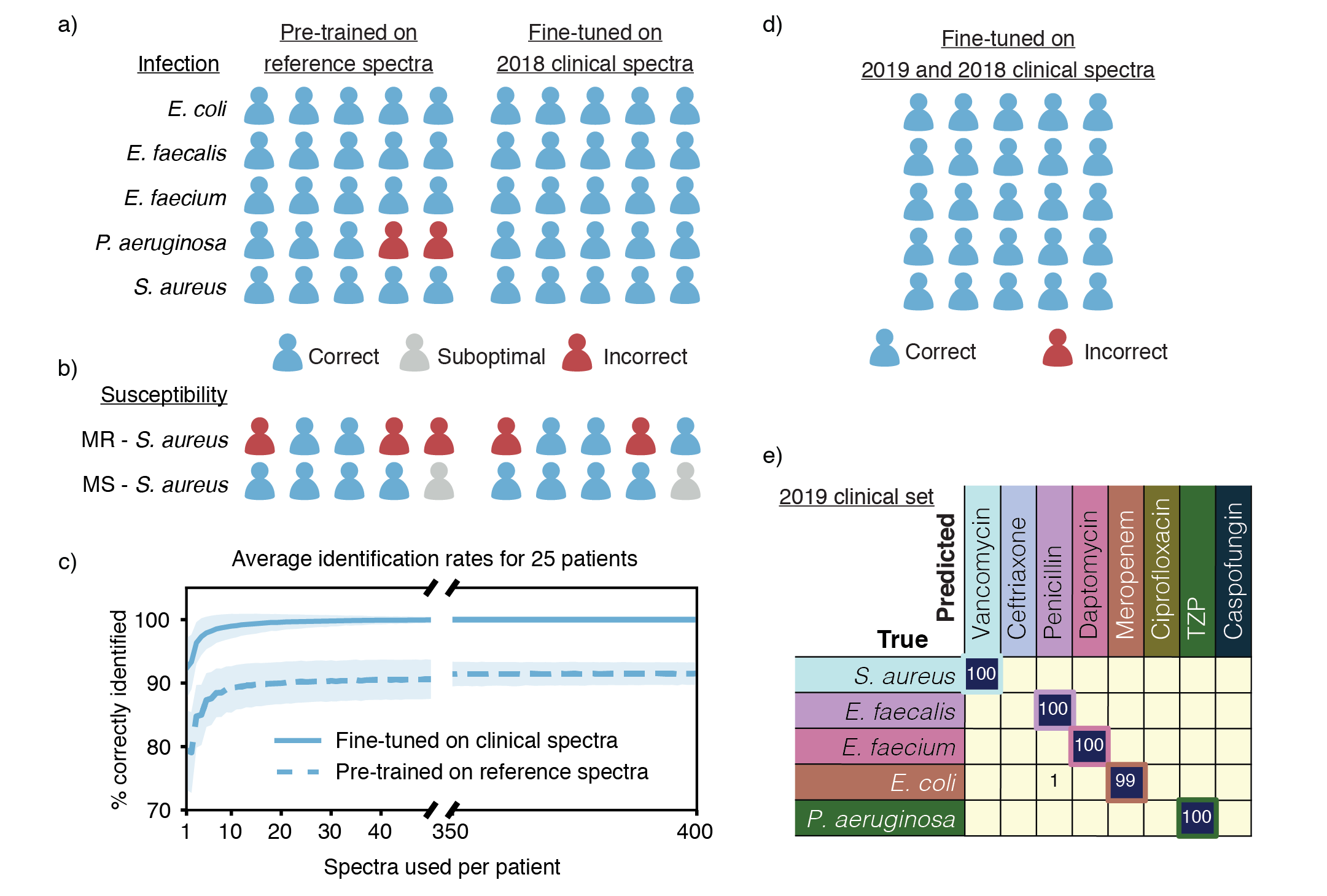}
\caption{Extension to clinical patient isolates. A CNN pre-trained on our reference dataset can be extended to classify clinical patient isolates and further improved by fine-tuning on a small number of clinical spectra.
a) 5 species of bacterial infections are tested, with 5 patients per infection type. Each patient is classified into one of 8 treatment classes where each species corresponds to a different treatment class. After fine-tuning, species identification accuracy improves from 89.0$\pm$3.6\% to 99.0$\pm$1.9\% ($\pm$ calculated as standard deviation across 10,000 sampling trials). 
b) Binary classification between MRSA and MSSA patient isolates is also performed, with an accuracy of 61.7$\pm$7.3\% that improves to 65.4$\pm$6.3\% after fine-tuning.
c) Dependence of average diagnosis rates for the fine-tuned model on the number of spectra used per patient. With just 10 spectra, the performance of the model reaches 99\% --- within 1\% difference of the performance with 400 spectra (100\%). Error bars are calculated as the standard deviation across 10,000 trials of random selections of n spectra, where n is the number of spectra used per patient.
d) We perform an additional test on a new clinical dataset gathered from an additional 25 patients with the same distribution across species as the first clinical dataset. We update the model that is pre-trained on the reference dataset and fine-tuned on the first clinical dataset by fine-tuning on the second clinical dataset using the same procedure.
e) Detailed breakdown by class for the second clinical dataset. Correct pairings between species and treatment group are outlined in the colored boxes. The rate of accurate identification is 99.7$\pm$1.1\%
}
\label{fig:clinical}
\end{figure}

\newpage
\clearpage
\section*{Methods}
\subsubsection*{Dataset}
The reference dataset consists of 30 bacterial and yeast isolates, including multiple isolates of Gram-negative and Gram-positive bacteria, as well as Candida species. We also include an isogenic pair of \textit{S. aureus} from the same strain, in which one variant contains the \textit{mecA} resistance gene for methicillin (MRSA) and the other does not (MSSA)\cite{Diep2008-op} (see Supplementary Table 1 for full isolate information). The reference training dataset consists of 2000 spectra each for the 30 reference isolates plus isogenic MSSA at 3 measurement times. The reference fine-tuning and test datasets each consist of 100 spectra for each of the 30 reference isolates.
The first clinical dataset consists of 30 patient isolates distributed across 5 species, with 400 spectra per isolate. The second clinical dataset consists of 25 patient isolates distributed across the same 5 species, with 100 spectra per isolate. Due to degradation in optical system efficiency, the measurement times for the reference fine-tuning and test and second clinical datasets were increased from 1 s to 2 s in order to keep SNR consistent across datasets.
Antibiotic susceptibility was performed by first genotypic testing for methicillin by detecting mecA using PCR (PMID: 19741081). Then phenotypic antimicrobial susceptibility testing was performed on the Microscan Walkaway instrument (Beckman Coulter, Brea, CA) and VITEK️ 2 (Biom\'erieux, Inc., Durham, NC).

\subsubsection*{Dataset variance}
For our datasets, we observe that intra-sample variance is high, as demonstrated by the pairwise spectral difference analysis summarized in Supplementary Figure 2. For 19 out of 30 isolates, spectra from at least one other isolate are more similar on average than spectra from the same isolate, on average. For example, when we rank isolates in order of similarity to \textit{E. faecalis} 2 (Supplementary Figure 2c), there are 8 other isolates where the average difference between a spectrum from \textit{E. faecalis} 2 and a spectrum from the other isolate is smaller than the average difference between two spectra from \textit{E. faecalis} 2. When intra-sample variance is high, a large number of spectra per sample may help to better represent the full data distribution and lead to higher predictive performance.

\subsubsection*{Sample preparation}
Bacterial isolates were cultured on blood agar plates each day before measurement. Plates were sealed with Parafilm and stored at 4\degree C for 20 minutes to 12 hours before sample preparation. Storage times varied to allow for multiple measurement times per day; however all other sample preparation conditions were kept consistent between samples. Differences in storage time were not found to result in spectral changes greater than spectral changes due to strain or isogenic differences. All clinical isolates were prepared in separate samples with consistent sample preparation conditions. Because test samples were prepared separately from samples used for training, we conclude that classifications are not due to batch effects such as differences in sample preparation. We prepared samples for measurement by suspending 0.6 mg of biomass from a single colony in 10 $\mu$L of sterile water (0.4 mg in 5 $\mu$L water for Gram-positive species) and drying 3 $\mu$L of the suspension on a gold-coated silica substrate (Figure 1a and b). Substrates were prepared by electron beam evaporation of 200 nm of gold onto microscope slides that were pre-cleaned using base piranha.
Samples were allowed to dry for 1 hour before measurement.

\subsubsection*{Raman measurements}
We measured Raman spectra across monolayer regions of the dried samples (Figure 1a) using the mapping mode of a Horiba LabRAM HR Evolution Raman microscope. 633 nm illumination at 13.17 mW was used with a 300 l/mm grating to generate spectra with 1.2 cm$^{-1}$ dispersion to maximize signal strength while minimizing background signal from autofluorescence. Wavenumber calibration was performed using a silicon sample. The 100X 0.9 NA objective lens (Olympus MPLAN) generates a diffraction-limited spot size, $\sim$1 $\mu$m in diameter. 
A 45x45 discrete spot map is taken with 3 $\mu$m spacing between spots to avoid overlap between spectra. The spectra are individually background corrected using a polynomial fit of order 5 using the subbackmod Matlab function available in the Biodata toolbox (see Supplementary Figure 1 for examples of raw and corrected spectra).
The majority of spectra are measured on true monolayers and arise from \~1 cell due to the diffraction-limited laser spot size, which is roughly the size of a bacteria cell. However, a small number of spectra may be taken over aggregates or multilayer regions. We exclude the spectra that are most likely to be non-monolayer measurements by ranking the spectra by signal intensity and discarding the 25 spectra with highest intensity, which includes all spectra with intensities greater than two standard deviations from the mean. We measured both monolayers and single cells, and found that monolayer measurements have SNRs of 2.5$\pm$0.7, similar to single-cell measurements (2.4$\pm$0.6), while allowing for the semi-automated generation of a large training dataset.
The spectral range between 381.98 and 1792.4 cm$^{-1}$ was used, and spectra were individually normalized to run from a minimum intensity of 0 and maximum intensity of 1 within this spectral range. SNR values are calculated by dividing the total intensity range by the intensity range over a 20-pixel wide window in a region where there is no Raman signal.

\subsubsection*{CNN architecture \& training details}
The CNN architecture is adapted from the Resnet architecture\cite{He2015-zl} that has been widely successful across a range of computer vision tasks.
It consists of an initial convolution layer followed by 6 residual layers and a final fully connected classification layer --- a block diagram can be seen in Figure 1.
The residual layers contain shortcut connections between the input and output of each residual block, allowing for better gradient propagation and stable training (refer to reference 37 for details).
Each residual layer contains 4 convolutional layers, so the total depth of the network is 26 layers.
The initial convolution layer has 64 convolutional filters, while each of the hidden layers has 100 filters.
These architecture hyperparameters were selected via grid search using one training and validation split on the isolate classification task. We also experimented with simple MLP (multi-layer perceptron) and CNN architectures but found that the Resnet-based architecture performed best.

We first train the network on the 30-isolate classification task, where the output of the CNN is a vector of probabilities across the 30 classes and the maximum probability is taken as the predicted class.
The binary MRSA/MSSA and binary isogenic MRSA/MSSA classifiers have the same architecture as the 30-isolate classifier, aside from the number of classes in the final classification layer.
We use the Adam optimizer\cite{Kingma2014-ni} across all experiments with learning rate 0.001, betas (0.5, 0.999), and batch size 10.
Classification accuracies are reported across 5 randomly selected train and validation splits.
We first pre-train the CNN on the reference training dataset, then fine-tune on the reference fine-tuning dataset to account for measurement changes due to degradation in optical system efficiency. For each of the 5 splits, we split the fine-tuning data into 90/10 train and validation splits, train the CNN on the train split, and use the accuracy on the validation split to perform model selection. We then evaluate and report the test accuracy on the test dataset which is gathered from independently cultured and prepared samples. The binary MRSA/MSSA classifier is trained and fine-tuned using the same procedure. The binary isogenic MRSA/MSSA classifier is trained using a similar procedure on data from a single measurement series.

All error values reported for tests on the reference dataset are standard deviation values across 5 splits.

While a high number of samples is good for ensuring dataset variation, deep learning approaches can still benefit from having a high number of examples per sample. When intra-sample variance is high, as we observe for our datasets, a large number of spectra per sample may better represent the full distribution and lead to higher predictive performance.

For the clinical isolates, we start by pre-training a CNN on the empiric treatment labels for the 30 reference isolates. We then use the following leave-one-patient-out cross-validation (LOOCV) strategy to fine-tune the parameters of the CNN. There are a total of 25 patient isolates across 5 species. In each of the 5 folds, we assign 1 patient in each species to the test set, 1 patient in each species to the validation set, and the remaining 3 patients in each species to the training (i.e., fine-tuning) set. We then use the clinical training set (consisting of isolates from 15 patients) to fine-tune the CNN parameters, and use accuracy on the validation set (5 patient isolates) to do model selection. The test accuracy for each fold is evaluated on the test set (5 patient isolates) using the method described below.

\subsubsection*{Clinical identification data analysis}
To reach an identification for patient isolates, 400 spectra are measured across a sample from each patient isolate. 10 of these spectra are chosen at random to be classified. The most common class out of the 10 spectral classifications is then chosen as the identification for each patient isolate, with ties broken randomly.  All error values reported for tests on the clinical dataset are standard deviations across 10,000 trials of random selections of 10 spectra, with an upper accuracy bound of 100\%. For the second clinical dataset, we perform the same procedure, except that we choose 10 out of 100 spectra for each patient isolate, and use a model that is both pre-trained on the reference dataset and fine-tuned on the first clinical dataset.

\subsubsection*{Baselines}
In all experiments where logistic regression (LR) and support vector machine (SVM) baselines were used, we first used PCA to reduce the input dimension from 1000 to 20 --- this hyperparameter was determined by plotting test accuracies for different settings on one training and validation split for the 30 isolate task and picking a value near where the test accuracy saturated. Using only the first 20 principal components not only decreases computation costs, but also increases accuracy by reducing the amount of noise in the data. For each fold of the cross validation procedure, we use grid search to choose the regularization hyperparameter for each model achieving the best validation accuracy and report the corresponding test accuracy. Using both the training and fine-tuning reference datasets to train the baseline models, LR and SVM achieve 57.5\% and 56.8\% on the 30-class task and 89.0\% and 88.3\% on the empiric treatment task, respectively. Using only the fine-tuning reference dataset, LR and SVM achieve 75.7\% and 74.9\% on the 30-class task and 93.3\% and 92.2\% on the empiric treatment task, respectively. The latter performance is higher because the baseline models do not benefit from additional training data as the CNN does, but rather benefit from training data the most closely matches the measurement conditions of the test data.

\subsubsection*{Two-sample test of sample means}
We use the Welch's two-sample $t$-test to test whether the differences in mean clinical accuracy for the CNN and the SVM and LR baselines were statistically significant. Welch's $t$-test is a  variation of the Student's $t$-test that is used when the two samples may have unequal variances.
In each case, we start by computing the pooled standard deviation as
\begin{equation}
\sigma = \sqrt{ \frac{(n_1 - 1) \sigma_1^2 + (n_2 - 1) \sigma_2^2}{n_1 + n_2 - 2} }.
\end{equation}
We then compute the standard error of the difference between the means as
\begin{equation}
\textrm{se} = \sigma \times \sqrt{\frac{1}{n_1} + \frac{1}{n_2}}.
\end{equation}
Finally, we can compute the test statistic as
\begin{equation}
t = \frac{\mu_1 - \mu_2}{\textrm{se}},
\end{equation}
and then compute the p-value using the corresponding Student's $t$-distribution. For our computations, $n_{\text{CNN}} = n_{\text{LR}} = n_{\text{SVM}} = 10000$, $\mu_{\text{CNN}} = 89.0$, $\mu_{\text{LR}} = 81.8$, $\mu_{\text{SVM}} = 82.9$, $\sigma_{\text{CNN}} = 3.6$, $\sigma_{\text{LR}} = 6.0$, and $\sigma_{\text{SVM}} = 5.9$. In comparing the CNN with LR, we computed a $t$-statistic of 102.9 and in comparing the CNN with SVM, we computed a $t$-statistic of 88.3. In both cases, we reject the null hypothesis that the means are equal at the 1e-6 p-level.

\subsubsection*{Data availability}
All data needed to replicate these results are available at https://github.com/csho33/bacteria-ID.

\subsubsection*{Code availability}
All code needed to replicate these results is available at https://github.com/csho33/bacteria-ID.

\subsubsection*{Biological materials availability}
Unique isolates are available from the authors upon reasonable request. 


\newpage
\clearpage
\setcounter{table}{0}
\setcounter{figure}{0}
\renewcommand{\figurename}{Supplementary Figure}
\renewcommand{\tablename}{Supplementary Table}

\begin{table} \centering \spacing{1}
\includegraphics[width=1\textwidth]{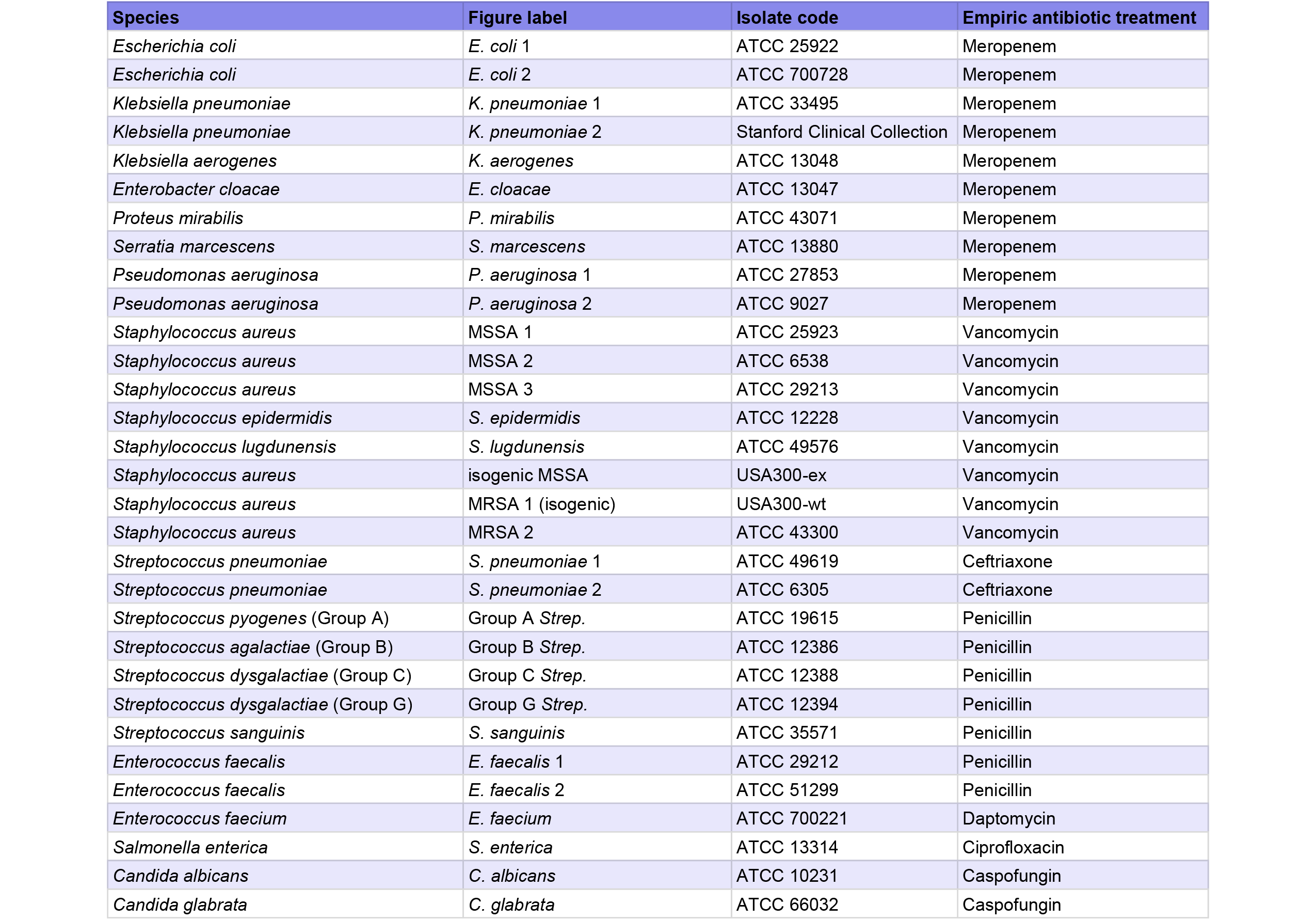}
\caption{Reference isolates. The empiric treatments are chosen by the authors of this paper specializing in infectious diseases from recommendations from Sanford Guide to Antimicrobial Therapy and trends in patient susceptibility profiles at the Stanford Hospital and the Veterans Affairs Palo Alto Health Care System \cite{Banaei_undated-ux,Nakasone2017-eb}. However, specific choices for each of the empiric species groups may be modified according to individual hospital susceptibility profiles.
}
\label{table:cell_lines}
\end{table}

\begin{table} \centering \spacing{1}
\includegraphics[width=1\textwidth]{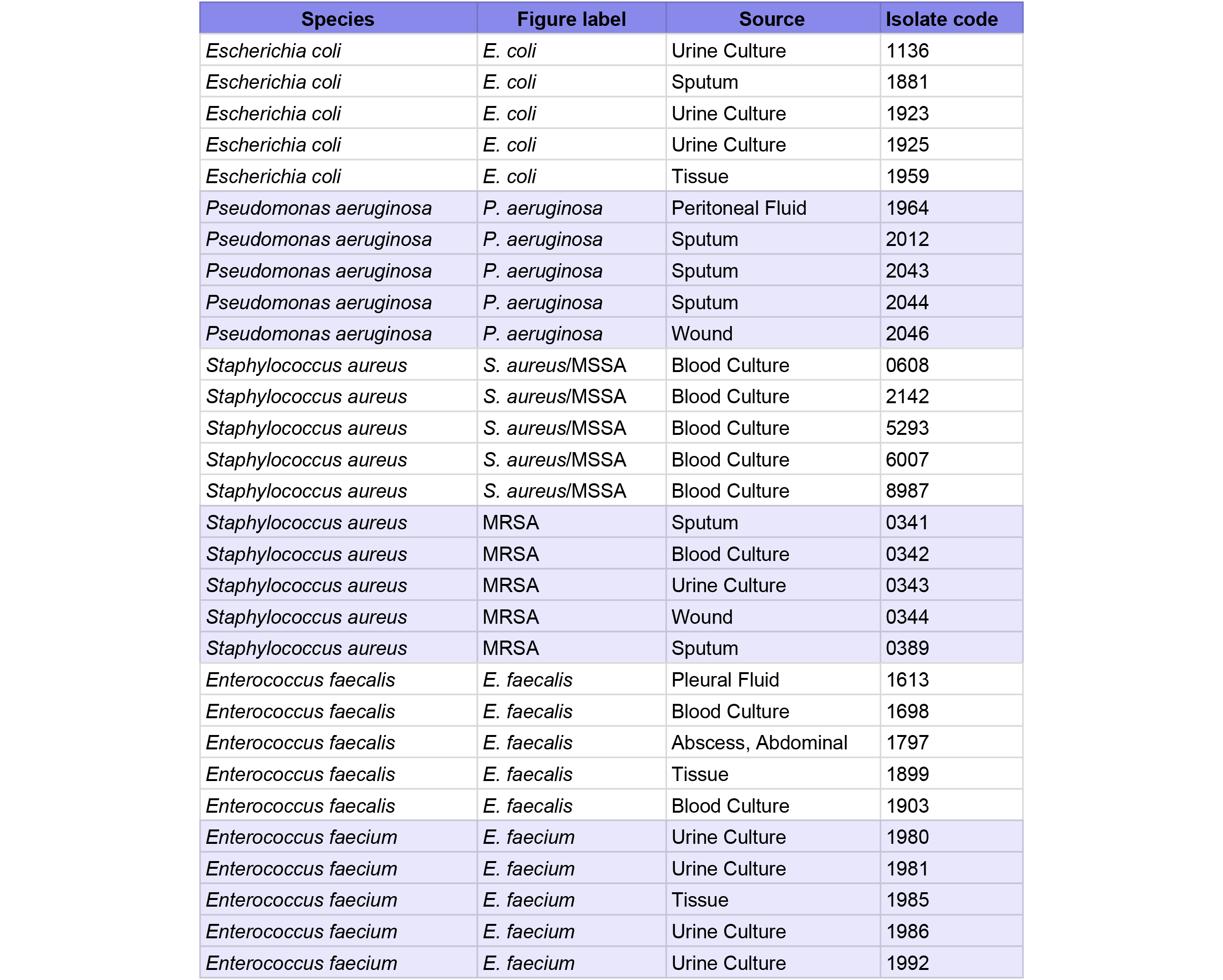}
\caption{Clinical isolates
}
\label{table:isolates}
\end{table}

\begin{figure} \centering \spacing{1}
\includegraphics[width=1\textwidth]{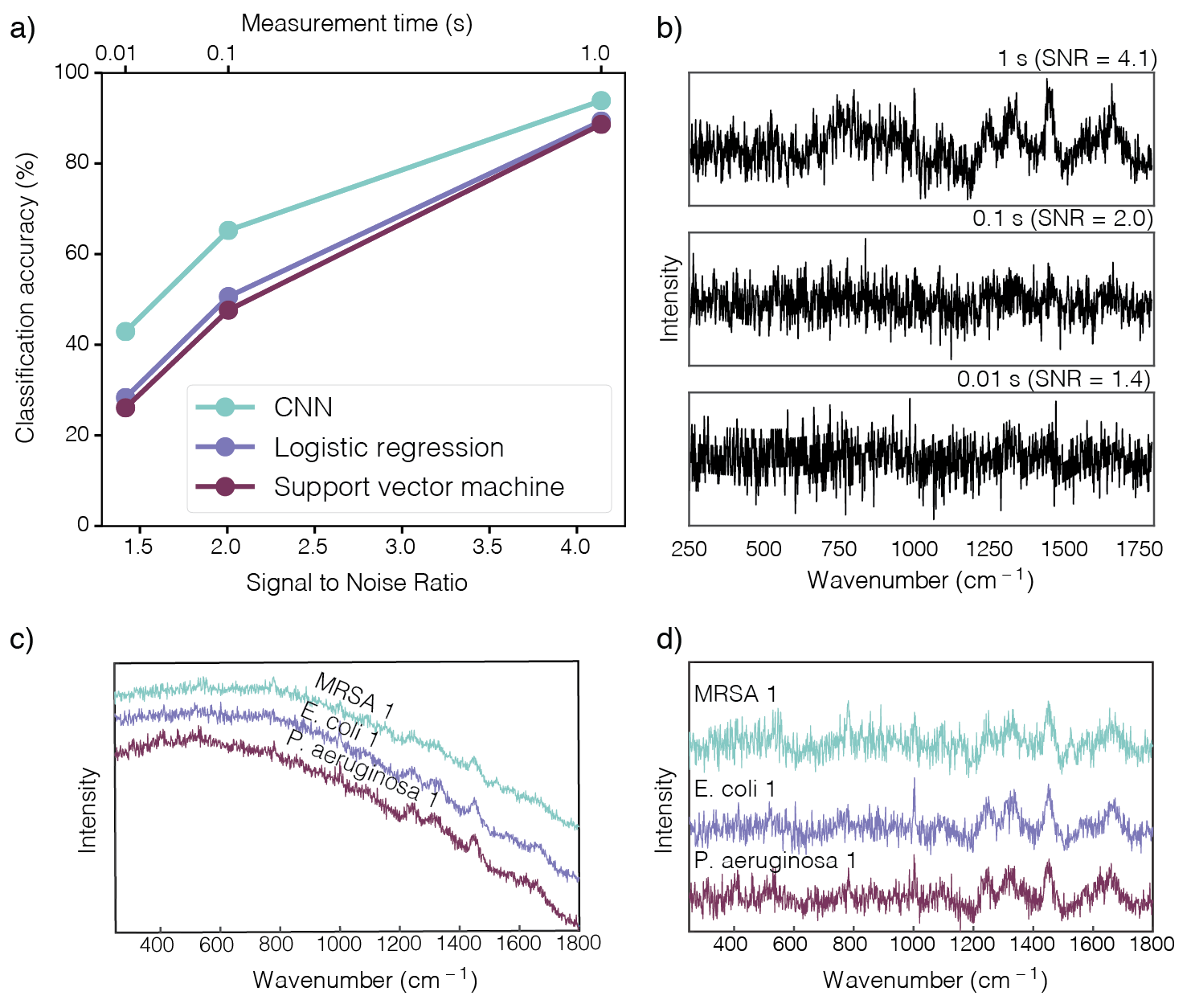}
\caption{a) Isolate-level classification accuracy increases with SNR. Under the measurement conditions used in this study, performance of the CNN is negatively affected by shorter measurement times. Further increase of SNR should saturate the performance of the CNN to a minimal baseline error rate. For this experiment, training, validation, and test sets are split between a single measurement series for each isolate.
b) Spectral examples (from \textit{E. coli} 1) for measurement times of 1 s, 0.1 s, and 0.01 s.
c) Raw spectra for MRSA 1, \textit{E. coli} 1, and \textit{P. aeruginosa} 1 for a measurement time of 1 s.
d) Spectra after background subtraction and normalization for a measurement time of 1 s. These are the direct inputs into our model.
}
\label{fig:snr}
\end{figure}

\begin{figure} \centering \spacing{1}
\includegraphics[width=1\textwidth]{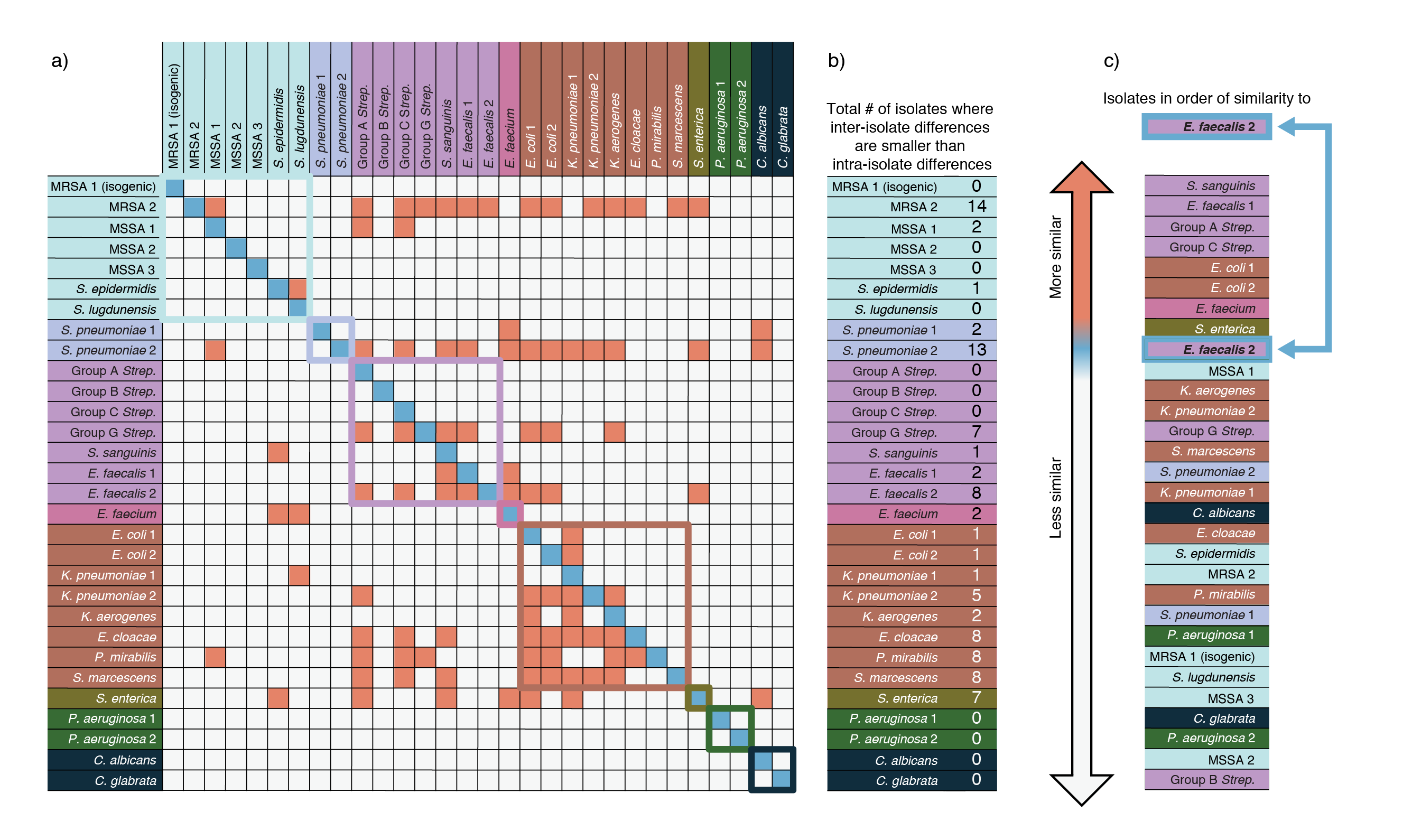}
\caption{Inter-isolate vs intra-isolate pairwise spectral differences. Average differences are calculated as the average L2 distance between pairs of spectra over 4 million (2000 x 2000) possible pairs. 
a) Intra-isolate distances (along the diagonal) are computed as the difference between two spectra from the same isolate, while inter-isolate distances (off-diagonals) are computed as the difference between one spectrum from the row isolate and one spectrum from the column isolate. For each row, red marks indicate isolates for which inter-isolate differences are smaller  than the average intra-isolate difference for the isolate in that row. Blue marks simply indicate the location of the diagonal for reference. For example, in the second row, the average distance between an MSSA 1 spectrum and an MRSA 1 spectrum is smaller than the average distance between two MRSA 1 spectra — in other words, MSSA 1 and MRSA 1 spectra are more similar (on average) than MRSA 1 spectra are to themselves (on average).  
b) For each isolate, we summarize the total number of more similar isolates. For 19 out of 30 isolates, spectra from at least one other isolate are more similar than spectra from the same isolate. 
c) Example sort by similarity for E. faecalis 2, demonstrating that spectra from 8 isolates are more similar on average to E. faecalis 2 than different spectra from E. faecalis itself, on average.
}
\label{fig:diff}
\end{figure}

\newpage
\clearpage
\begin{figure} \centering \spacing{1}
\includegraphics[width=0.5\textwidth]{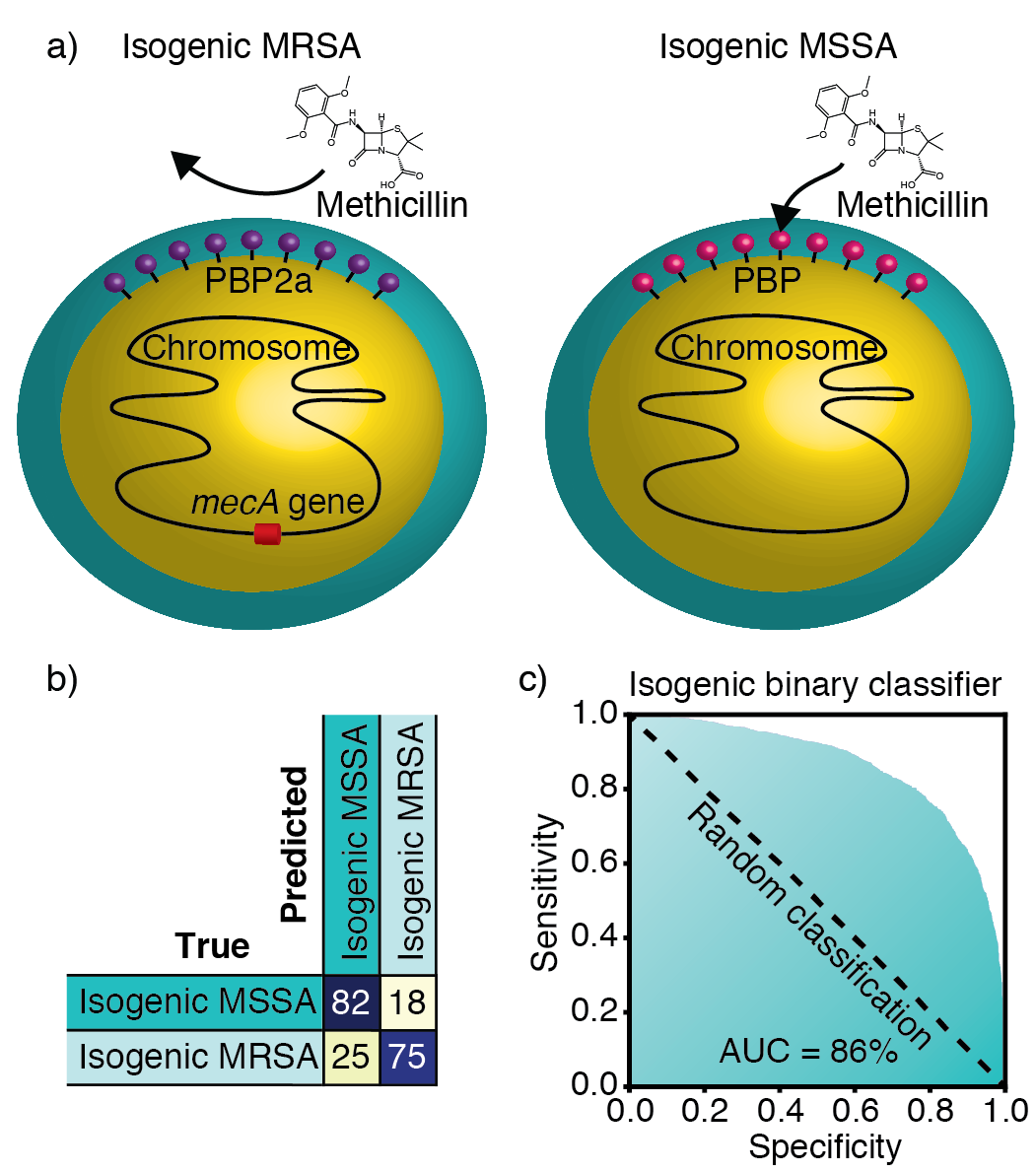}
\caption{Isogenic MRSA/MSSA classifier.
a) Sensitivity to antibiotic resistance alone with all other factors held constant can be tested using an isogenic pair of \textit{S. aureus}, meaning that the two are genetically identical aside from the deletion of the \textit{mecA} gene which confers methicillin resistance\cite{Diep2008-op}. The expression of \textit{mecA} results in replacement of Penicillin Binding Proteins (PBPs) with PBP2a, which has a low binding affinity for methicillin.
b) A binary classifier is trained to distinguish between MRSA 1 and its isogenic variant, achieving 78.5$\pm$0.6\% accuracy. For this experiment, training, validation, and test sets are split between a single measurement series for each isolate. These results are a first step in ongoing work aiming to understand whether isogenic pairs can be distinguished by their Raman spectra. Because the measured spectral differences are so small between isogenic pairs, we expect that true signal differences may be confounded by experimental factors including minute differences in sample drying time, incubation time, and sample positioning. These confounding factors would need to be carefully controlled for in future experiments where training, validation, and test sets are split between independently cultured and prepared samples. c) The ROC shows sensitivities and specificities significantly higher than random classification, with an AUC of 86.1\%.
}
\label{fig:iso}
\end{figure}

\begin{figure} \centering \spacing{1}
\includegraphics[width=1\textwidth]{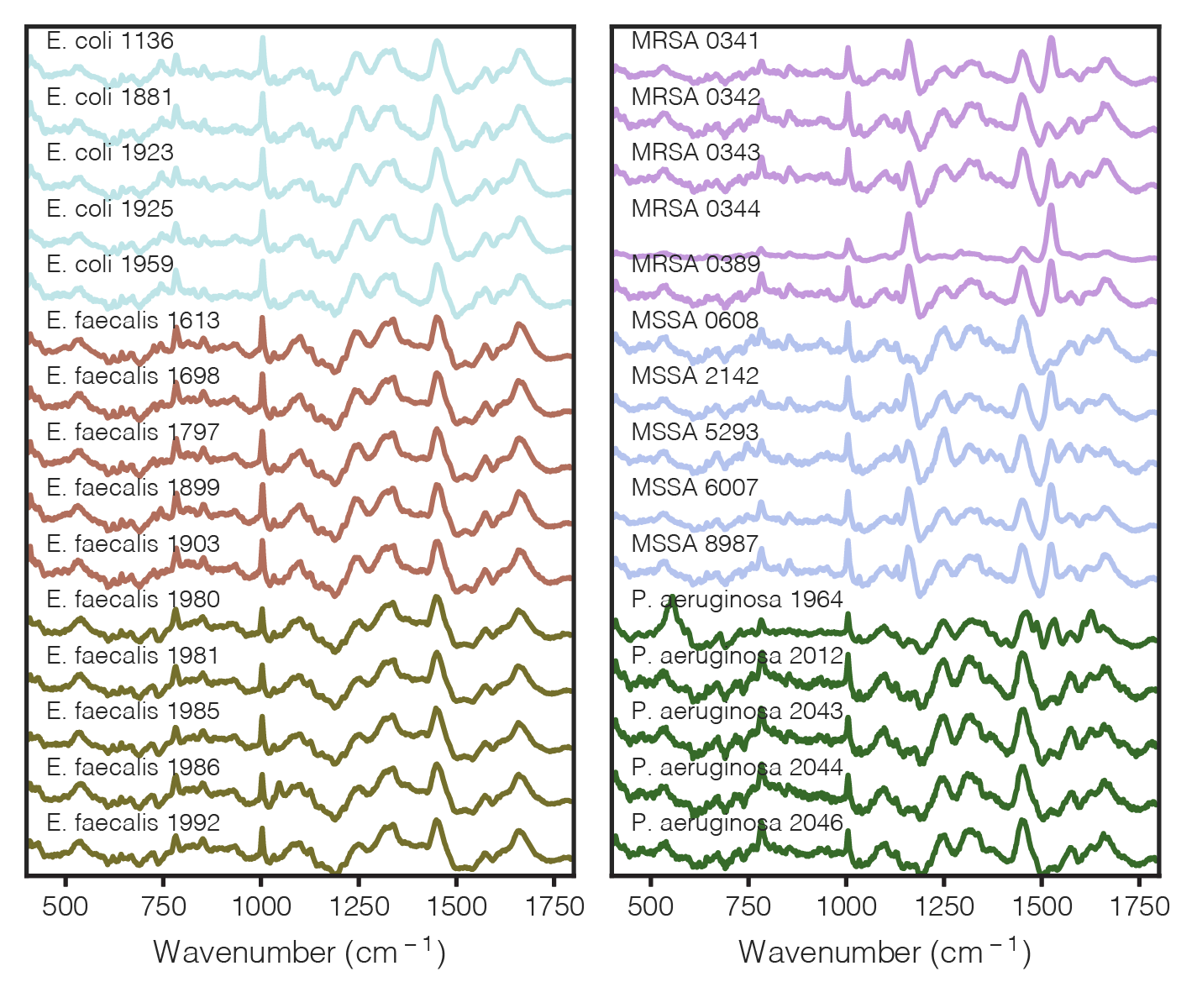}
\caption{Spectra for individual patient isolates, averaged across the full 400 spectra dataset for each patient.}
\label{fig:patients}
\end{figure}

\begin{figure} \centering \spacing{1}
\includegraphics[width=0.9\textwidth]{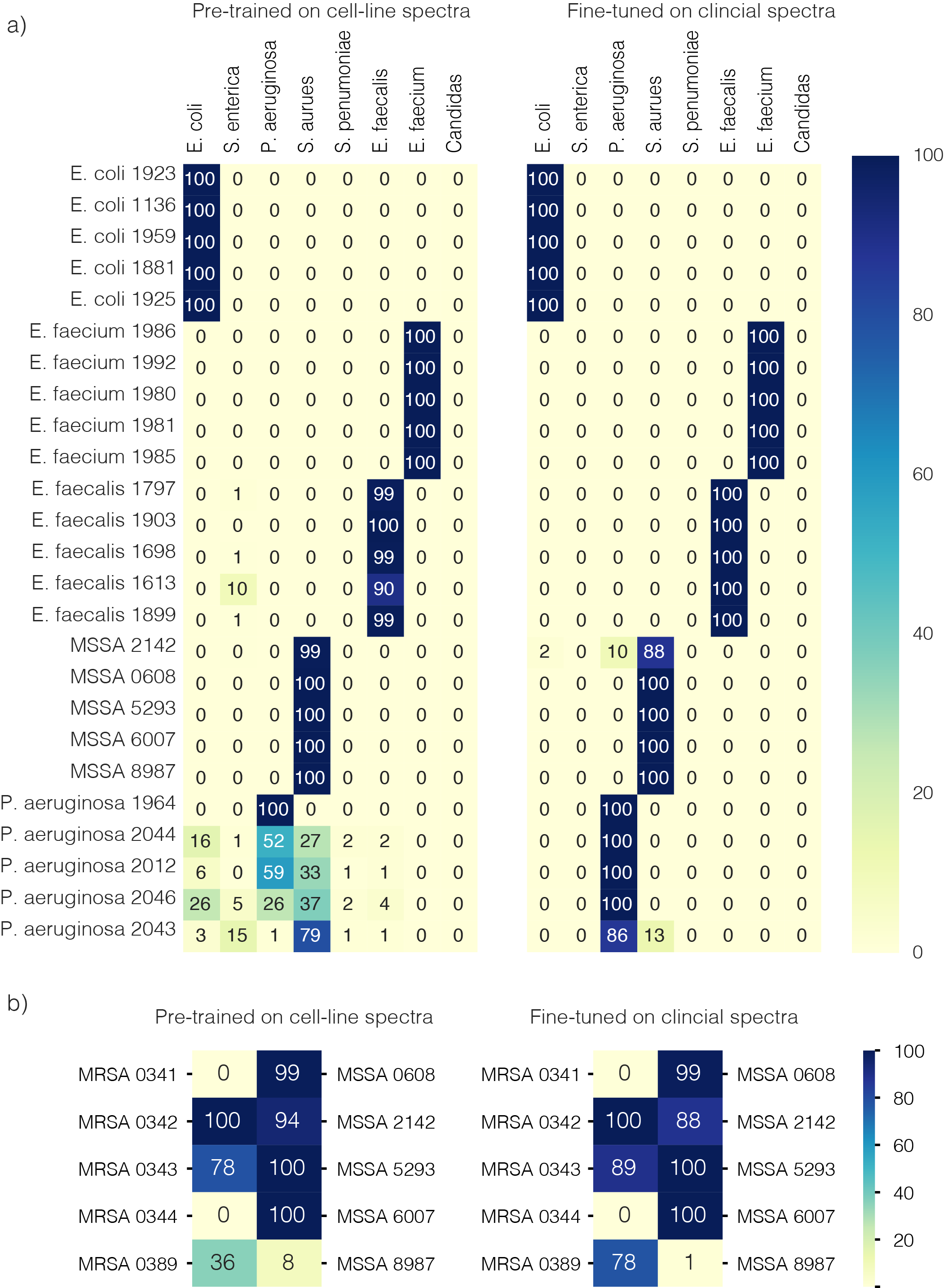}
\caption{a) Classification results for each patient isolate. Element (i, j) represents the percentage out of 10,000 trials in which species j is predicted by the CNN for patient i. 
b)Classification results for each MRSA/MSSA patient isolate. Heatmap represents the percentage out of 10,000 trials in which the binary CNN accurately identifies whether the isolate is MRSA or MSSA. 10 spectra per isolate are used for both fine tuning and identification.
}
\label{fig:cm_clinical}
\end{figure}

\begin{figure} \centering \spacing{1}
\includegraphics[width=0.5\textwidth]{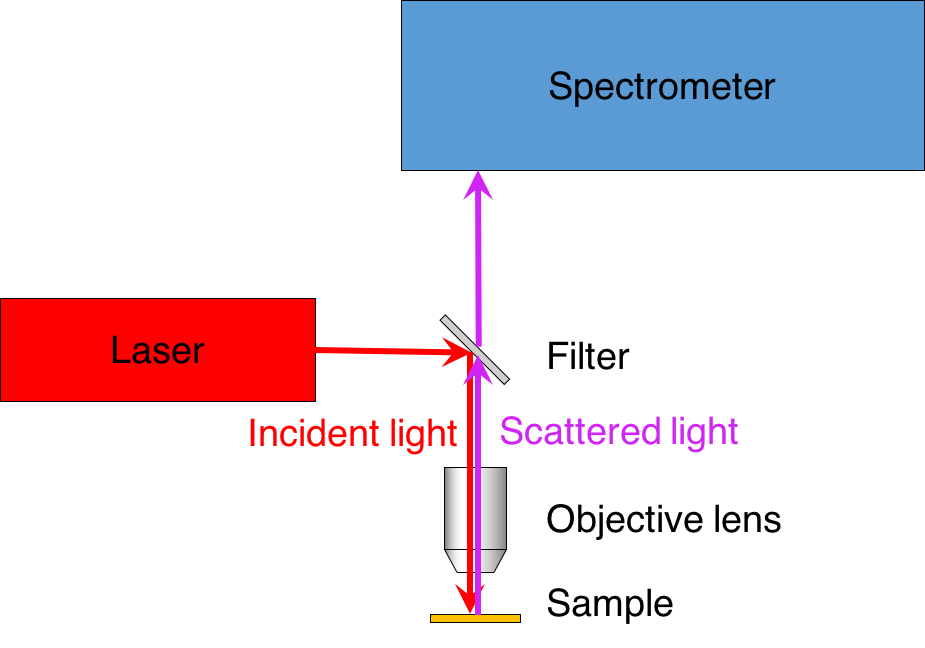}
\caption{Experimental schematic of the Horiba Labram Raman spectrometer.
}
\label{fig:labram}
\end{figure}

\begin{figure} \centering \spacing{1}
\includegraphics[width=0.6\textwidth]{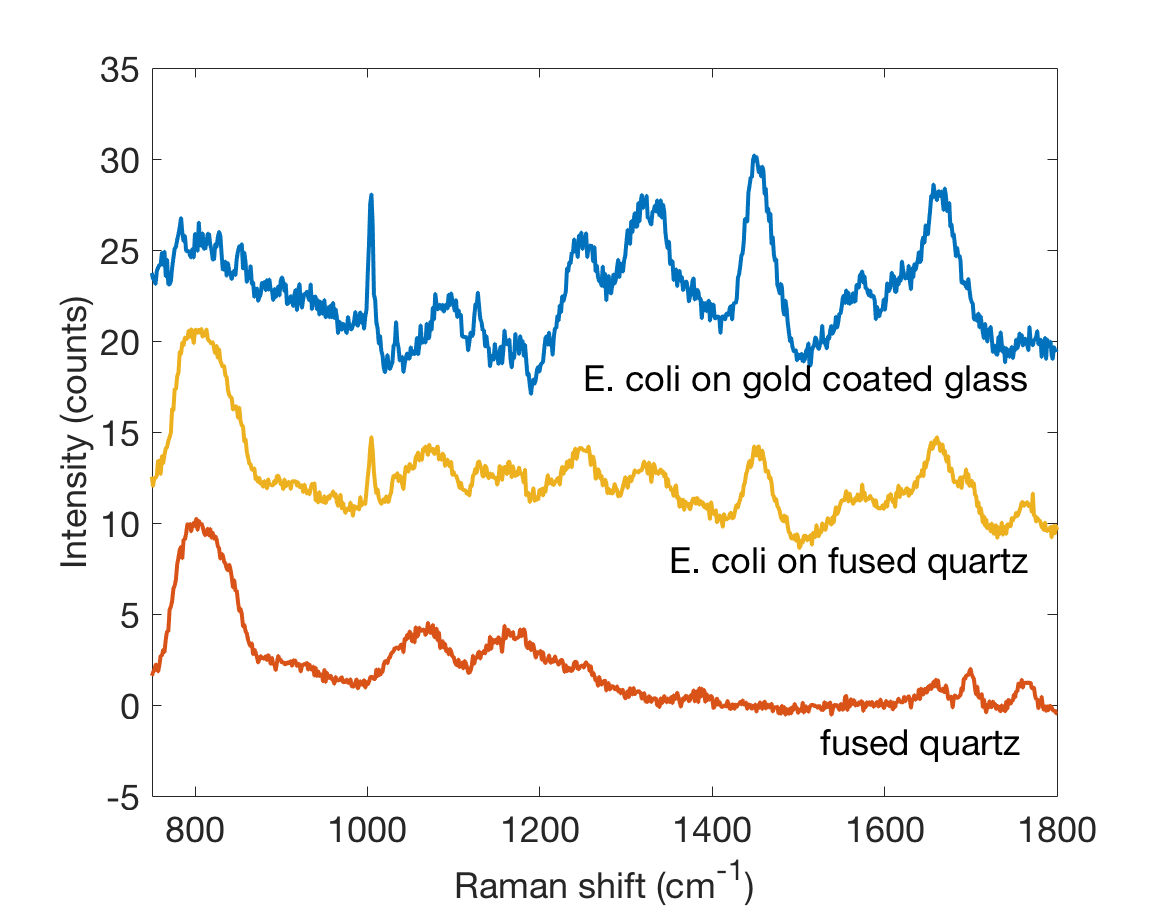}
\caption{Comparison of signal intensity on reflective and non-reflective substrates. We find that the signal intensity and SNR of our measurements on gold substrates is ~2X the the signal intensity and SNR of measurements on glass substrates. Because quartz is transparent at visible wavelengths and gold is reflective, it is more likely that this 2X enhancement is due to the reflection of forward-scattered photons rather than a SERS enhancement. These measurements were taken with the same measurement conditions as our datasets, but consist of 100 1s accumulations to help visualize the spectral shape with less noise.
}
\label{fig:substrate}
\end{figure}

\newpage
\clearpage
\begin{figure} \centering \spacing{1}
\includegraphics[width=1\textwidth]{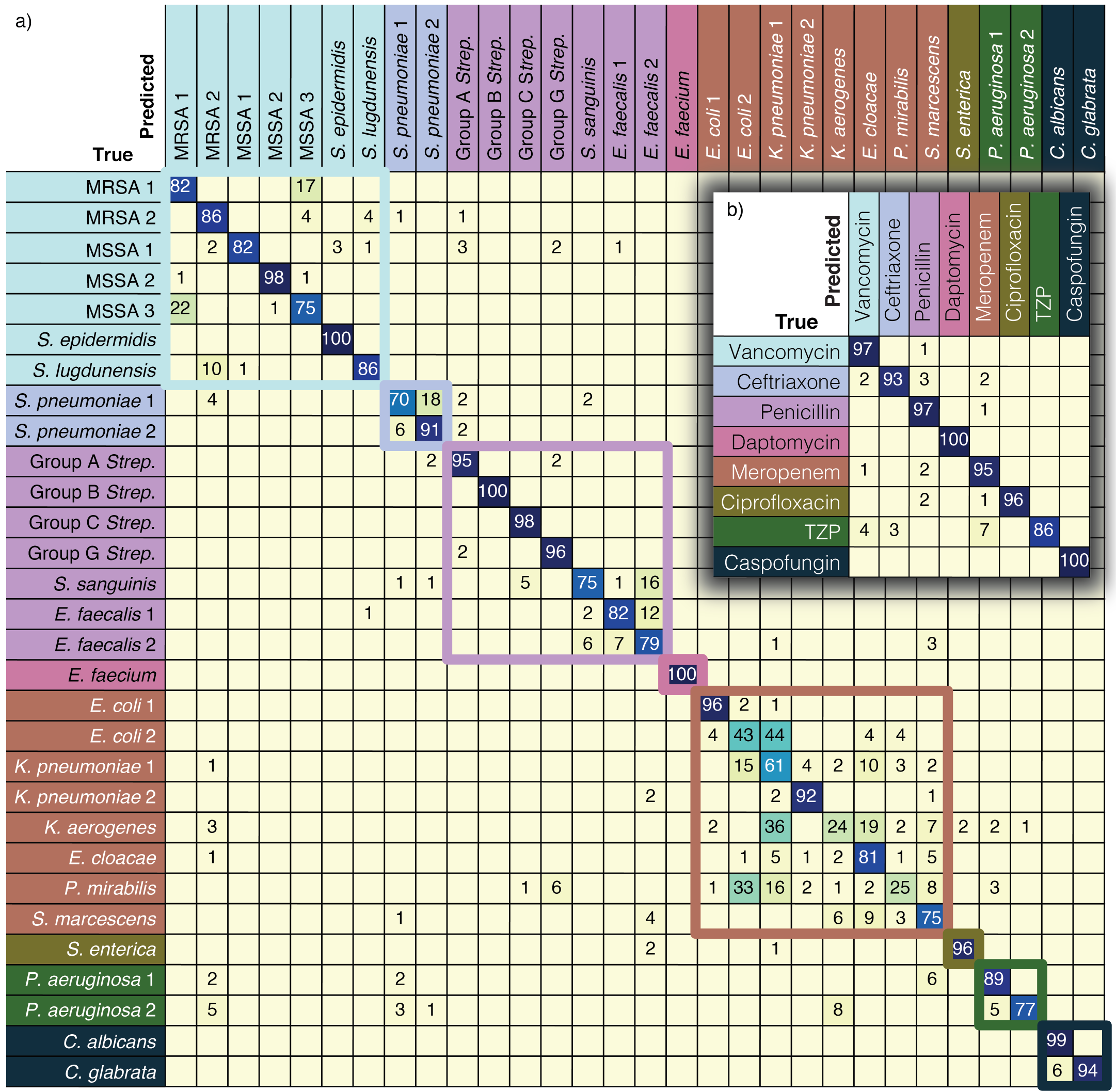}
\caption{CNN performance breakdown by class with test and fine-tune datasets swapped. The trained CNN classifies 30 bacterial and yeast isolates with isolate-level accuracy of 81.6$\pm$0.6\% and antibiotic grouping-level accuracy of 95.9$\pm$0.6\%.
a) Confusion matrix for 30 strain classes. Entry (i, j) represents the percentage out of 100 test spectra that are predicted by the CNN as class j given a ground truth of class i; entries along the diagonal represent the accuracies for each class. Misclassifications are mostly within antibiotic groupings, indicated by colored boxes, and thus do not affect the treatment outcome.
b) Predictions can be combined into antibiotic groupings to estimate treatment accuracy. TZP = piperacillin-tazobactam. All values below 0.5\% are not shown.
}
\label{fig:cm_swap}
\end{figure}

\begin{figure} \centering \spacing{1}
\includegraphics[width=0.3\textwidth]{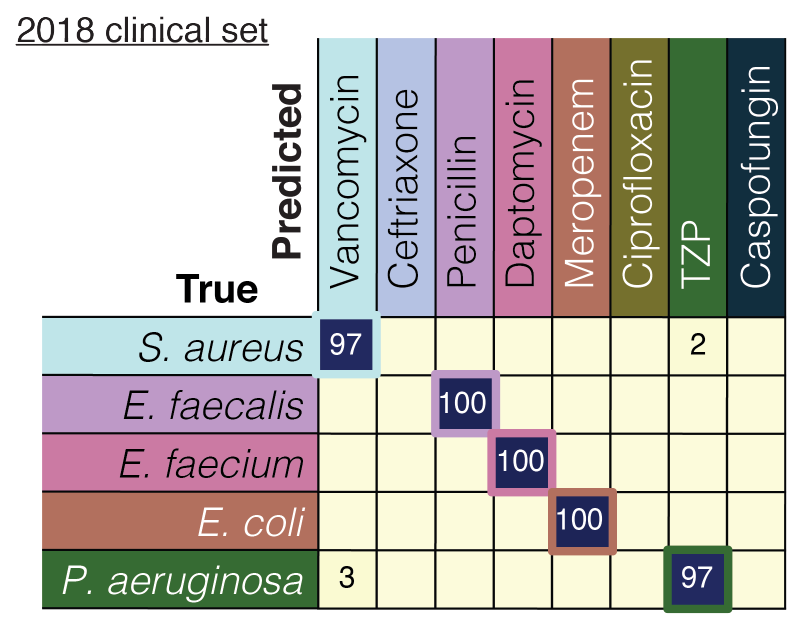}
\caption{Detailed breakdown by class for the first clinical dataset. Each patient is classified into one of 8 treatment classes where each species corresponds to a different treatment class. Correct pairings between species and treatment group are outlined in the colored boxes. The rate of accurate identification is 99.0$\pm$1.9\%.
}
\label{fig:cm_clinical_2018}
\end{figure}

\begin{figure} \centering \spacing{1}
\includegraphics[width=0.6\textwidth]{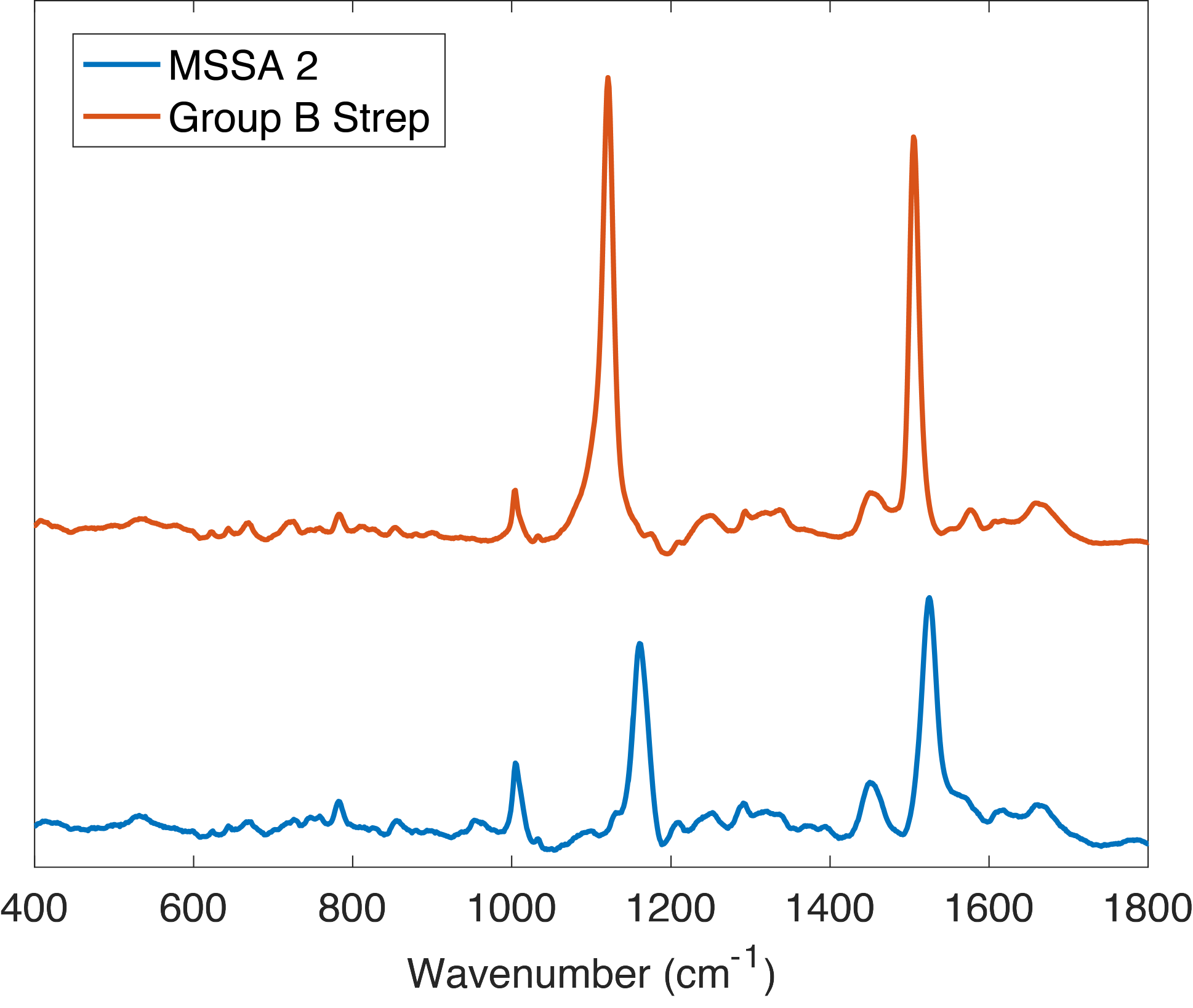}
\caption{The spectra of MSSA 2 and Group B \textit{Strep.} demonstrate resonant Raman effects from chromophores (e.g. carotenoids or cytochromes), resulting in enhanced Raman peaks around 1005 $cm^{-1}$, 1121-1162 $cm^{-1}$, and 1505-1525 $cm^{-1}$ \cite{Lorenz2017-lt}.
}
\label{fig:resonant}
\end{figure}

\setcounter{figure}{0}
\renewcommand{\figurename}{Supplementary Note}
\begin{figure} \centering \spacing{1}
\includegraphics[width=1\textwidth]{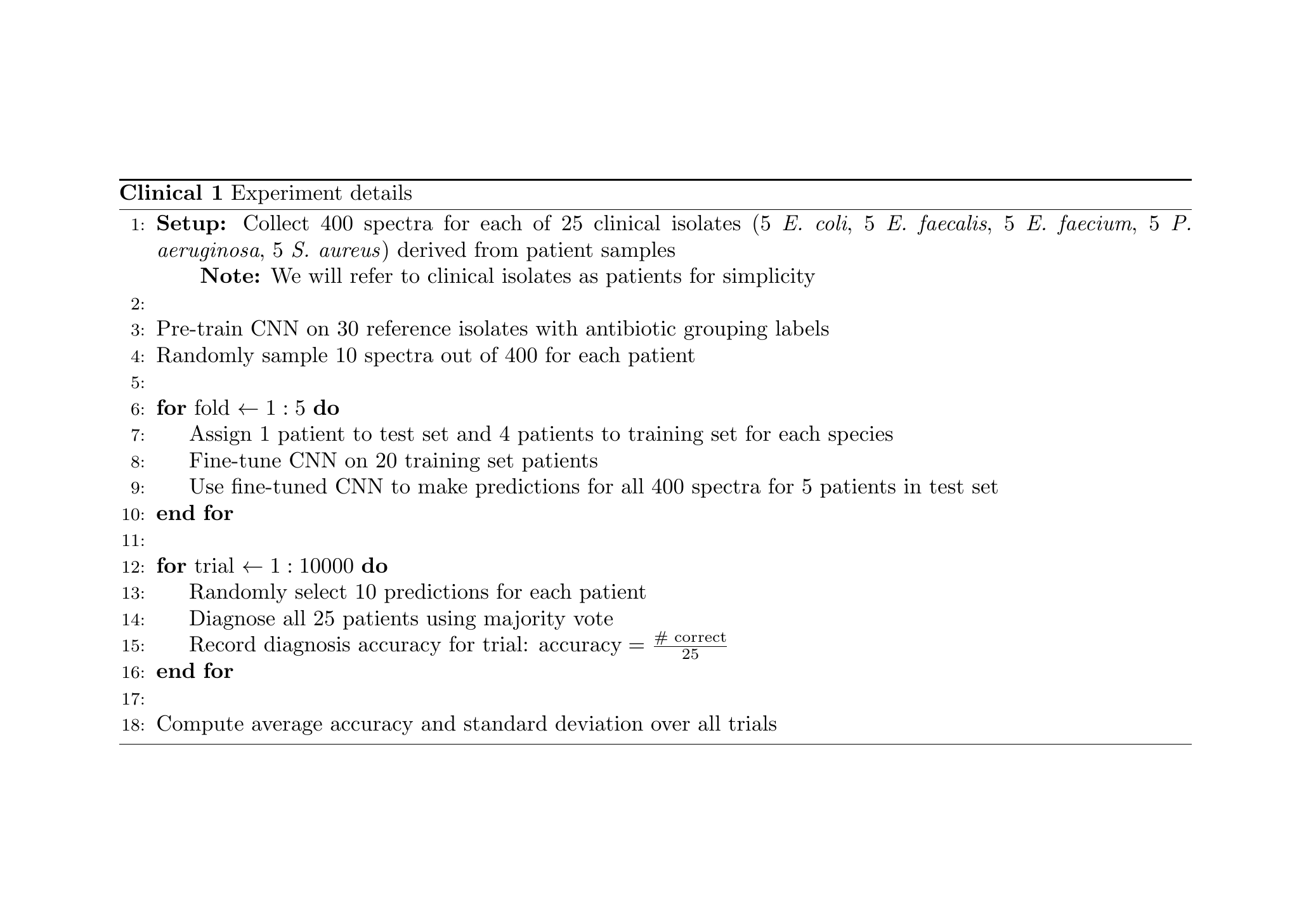}
\caption{Pseudocode for fine-tuning and identification of clinical spectra.
}
\label{fig:pseudocode}
\end{figure}


\newpage
\clearpage
\pagebreak
\baselineskip12pt
\bibliography{scibib}

\begin{thebibliography}{10}
\expandafter\ifx\csname url\endcsname\relax
  \def\url#1{\texttt{#1}}\fi
\expandafter\ifx\csname urlprefix\endcsname\relax\def\urlprefix{URL }\fi
\providecommand{\bibinfo}[2]{#2}
\providecommand{\eprint}[2][]{\url{#2}}

\bibitem{Fleischmann2016-uc}
\bibinfo{author}{Fleischmann, C.} \emph{et~al.}
\newblock \bibinfo{title}{Assessment of global incidence and mortality of
  hospital-treated sepsis. current estimates and limitations}.
\newblock \emph{\bibinfo{journal}{Am. J. Respir. Crit. Care Med.}}
  \textbf{\bibinfo{volume}{193}}, \bibinfo{pages}{259--272}
  (\bibinfo{year}{2016}).

\bibitem{DeAntonio2016-en}
\bibinfo{author}{DeAntonio, R.}, \bibinfo{author}{Yarzabal, J.-P.},
  \bibinfo{author}{Cruz, J.~P.}, \bibinfo{author}{Schmidt, J.~E.} \&
  \bibinfo{author}{Kleijnen, J.}
\newblock \bibinfo{title}{Epidemiology of community-acquired pneumonia and
  implications for vaccination of children living in developing and newly
  industrialized countries: A systematic literature review}.
\newblock \emph{\bibinfo{journal}{Hum. Vaccin. Immunother.}}
  \textbf{\bibinfo{volume}{12}}, \bibinfo{pages}{2422--2440}
  (\bibinfo{year}{2016}).

\bibitem{Torio2016-vz}
\bibinfo{author}{Torio, C.~M.} \& \bibinfo{author}{Moore, B.~J.}
\newblock \bibinfo{title}{National inpatient hospital costs: The most expensive
  conditions by payer, 2013}.
\newblock \bibinfo{type}{Tech. Rep.} \bibinfo{number}{HCUP Statistical Brief
  \#204.}, \bibinfo{institution}{Agency for Healthcare Research and Quality}
  (\bibinfo{year}{2016}).

\bibitem{Dellinger2013-xq}
\bibinfo{author}{Dellinger, R.~P.} \emph{et~al.}
\newblock \bibinfo{title}{Surviving sepsis campaign: international guidelines
  for management of severe sepsis and septic shock: 2012}.
\newblock \emph{\bibinfo{journal}{Crit. Care Med.}}
  \textbf{\bibinfo{volume}{41}}, \bibinfo{pages}{580--637}
  (\bibinfo{year}{2013}).

\bibitem{Chaudhuri2008-ac}
\bibinfo{author}{Chaudhuri, A.} \emph{et~al.}
\newblock \bibinfo{title}{{EFNS} guideline on the management of
  community-acquired bacterial meningitis: report of an {EFNS} task force on
  acute bacterial meningitis in older children and adults}.
\newblock \emph{\bibinfo{journal}{Eur. J. Neurol.}}
  \textbf{\bibinfo{volume}{15}}, \bibinfo{pages}{649--659}
  (\bibinfo{year}{2008}).

\bibitem{American_Thoracic_Society2005-xy}
\bibinfo{author}{{American Thoracic Society}} \& \bibinfo{author}{{Infectious
  Diseases Society of America}}.
\newblock \bibinfo{title}{Guidelines for the management of adults with
  hospital-acquired, ventilator-associated, and healthcare-associated
  pneumonia}.
\newblock \emph{\bibinfo{journal}{Am. J. Respir. Crit. Care Med.}}
  \textbf{\bibinfo{volume}{171}}, \bibinfo{pages}{388--416}
  (\bibinfo{year}{2005}).

\bibitem{Fleming-Dutra2016-oq}
\bibinfo{author}{Fleming-Dutra, K.~E.} \emph{et~al.}
\newblock \bibinfo{title}{Prevalence of inappropriate antibiotic prescriptions
  among {US} ambulatory care visits, 2010-2011}.
\newblock \emph{\bibinfo{journal}{JAMA}} \textbf{\bibinfo{volume}{315}},
  \bibinfo{pages}{1864--1873} (\bibinfo{year}{2016}).

\bibitem{Butler2016-lb}
\bibinfo{author}{Butler, H.~J.} \emph{et~al.}
\newblock \bibinfo{title}{Using raman spectroscopy to characterize biological
  materials}.
\newblock \emph{\bibinfo{journal}{Nat. Protoc.}} \textbf{\bibinfo{volume}{11}},
  \bibinfo{pages}{664--687} (\bibinfo{year}{2016}).

\bibitem{Stockel2016-bg}
\bibinfo{author}{St{\"o}ckel, S.}, \bibinfo{author}{Kirchhoff, J.},
  \bibinfo{author}{Neugebauer, U.}, \bibinfo{author}{R{\"o}sch, P.} \&
  \bibinfo{author}{Popp, J.}
\newblock \bibinfo{title}{The application of raman spectroscopy for the
  detection and identification of microorganisms}.
\newblock \emph{\bibinfo{journal}{J. Raman Spectrosc.}}
  \textbf{\bibinfo{volume}{47}}, \bibinfo{pages}{89--109}
  (\bibinfo{year}{2016}).

\bibitem{Kloss2013-vr}
\bibinfo{author}{Kloss, S.} \emph{et~al.}
\newblock \bibinfo{title}{Culture independent raman spectroscopic
  identification of urinary tract infection pathogens: a proof of principle
  study}.
\newblock \emph{\bibinfo{journal}{Anal. Chem.}} \textbf{\bibinfo{volume}{85}},
  \bibinfo{pages}{9610--9616} (\bibinfo{year}{2013}).

\bibitem{Boardman2016-tp}
\bibinfo{author}{Boardman, A.~K.} \emph{et~al.}
\newblock \bibinfo{title}{Rapid detection of bacteria from blood with
  {Surface-Enhanced} raman spectroscopy}.
\newblock \emph{\bibinfo{journal}{Anal. Chem.}} \textbf{\bibinfo{volume}{88}},
  \bibinfo{pages}{8026--8035} (\bibinfo{year}{2016}).

\bibitem{Schmid2009-xt}
\bibinfo{author}{Schmid, U.} \emph{et~al.}
\newblock \bibinfo{title}{Gaussian mixture discriminant analysis for the
  single-cell differentiation of bacteria using micro-raman spectroscopy}.
\newblock \emph{\bibinfo{journal}{Chemometrics Intellig. Lab. Syst.}}
  \textbf{\bibinfo{volume}{96}}, \bibinfo{pages}{159--171}
  (\bibinfo{year}{2009}).

\bibitem{Munchberg2014-ee}
\bibinfo{author}{M{\"u}nchberg, U.}, \bibinfo{author}{R{\"o}sch, P.},
  \bibinfo{author}{Bauer, M.} \& \bibinfo{author}{Popp, J.}
\newblock \bibinfo{title}{Raman spectroscopic identification of single
  bacterial cells under antibiotic influence}.
\newblock \emph{\bibinfo{journal}{Anal. Bioanal. Chem.}}
  \textbf{\bibinfo{volume}{406}}, \bibinfo{pages}{3041--3050}
  (\bibinfo{year}{2014}).

\bibitem{Novelli-Rousseau2018-zo}
\bibinfo{author}{Novelli-Rousseau, A.} \emph{et~al.}
\newblock \bibinfo{title}{Culture-free antibiotic-susceptibility determination
  from single-bacterium raman spectra}.
\newblock \emph{\bibinfo{journal}{Sci. Rep.}} \textbf{\bibinfo{volume}{8}},
  \bibinfo{pages}{3957} (\bibinfo{year}{2018}).

\bibitem{Liu2016-vo}
\bibinfo{author}{Liu, C.-Y.} \emph{et~al.}
\newblock \bibinfo{title}{Rapid bacterial antibiotic susceptibility test based
  on simple surface-enhanced raman spectroscopic biomarkers}.
\newblock \emph{\bibinfo{journal}{Sci. Rep.}} \textbf{\bibinfo{volume}{6}},
  \bibinfo{pages}{23375} (\bibinfo{year}{2016}).

\bibitem{Lu2013-sd}
\bibinfo{author}{Lu, X.} \emph{et~al.}
\newblock \bibinfo{title}{Detecting and tracking nosocomial
  methicillin-resistant staphylococcus aureus using a microfluidic {SERS}
  biosensor}.
\newblock \emph{\bibinfo{journal}{Anal. Chem.}} \textbf{\bibinfo{volume}{85}},
  \bibinfo{pages}{2320--2327} (\bibinfo{year}{2013}).

\bibitem{Germond2018-ju}
\bibinfo{author}{Germond, A.} \emph{et~al.}
\newblock \bibinfo{title}{Raman spectral signature reflects transcriptomic
  features of antibiotic resistance in escherichia coli}.
\newblock \emph{\bibinfo{journal}{Communications Biology}}
  \textbf{\bibinfo{volume}{1}}, \bibinfo{pages}{85} (\bibinfo{year}{2018}).

\bibitem{Ayala2018-ie}
\bibinfo{author}{Ayala, O.~D.} \emph{et~al.}
\newblock \bibinfo{title}{{Drug-Resistant} staphylococcus aureus strains reveal
  distinct biochemical features with raman microspectroscopy}.
\newblock \emph{\bibinfo{journal}{ACS Infect Dis}}
  \textbf{\bibinfo{volume}{4}}, \bibinfo{pages}{1197--1210}
  (\bibinfo{year}{2018}).

\bibitem{Kirchhoff2018-ho}
\bibinfo{author}{Kirchhoff, J.} \emph{et~al.}
\newblock \bibinfo{title}{Simple ciprofloxacin resistance test and
  determination of minimal inhibitory concentration within 2 h using raman
  spectroscopy}.
\newblock \emph{\bibinfo{journal}{Anal. Chem.}} \textbf{\bibinfo{volume}{90}},
  \bibinfo{pages}{1811--1818} (\bibinfo{year}{2018}).

\bibitem{Vincent2009-kq}
\bibinfo{author}{Vincent, J.-L.} \emph{et~al.}
\newblock \bibinfo{title}{International study of the prevalence and outcomes of
  infection in intensive care units}.
\newblock \emph{\bibinfo{journal}{JAMA}} \textbf{\bibinfo{volume}{302}},
  \bibinfo{pages}{2323--2329} (\bibinfo{year}{2009}).

\bibitem{Krizhevsky2012-ke}
\bibinfo{author}{Krizhevsky, A.}, \bibinfo{author}{Sutskever, I.} \&
  \bibinfo{author}{Hinton, G.~E.}
\newblock \bibinfo{title}{{ImageNet} classification with deep convolutional
  neural networks}.
\newblock In \bibinfo{editor}{Pereira, F.}, \bibinfo{editor}{Burges, C. J.~C.},
  \bibinfo{editor}{Bottou, L.} \& \bibinfo{editor}{Weinberger, K.~Q.} (eds.)
  \emph{\bibinfo{booktitle}{Advances in Neural Information Processing Systems
  25}}, \bibinfo{pages}{1097--1105} (\bibinfo{publisher}{Curran Associates,
  Inc.}, \bibinfo{year}{2012}).

\bibitem{Mnih2014-ei}
\bibinfo{author}{Mnih, V.}, \bibinfo{author}{Heess, N.},
  \bibinfo{author}{Graves, A.} \& \bibinfo{author}{Kavukcuoglu, K.}
\newblock \bibinfo{title}{Recurrent models of visual attention}.
\newblock In \bibinfo{editor}{Ghahramani, Z.}, \bibinfo{editor}{Welling, M.},
  \bibinfo{editor}{Cortes, C.}, \bibinfo{editor}{Lawrence, N.~D.} \&
  \bibinfo{editor}{Weinberger, K.~Q.} (eds.) \emph{\bibinfo{booktitle}{Advances
  in Neural Information Processing Systems 27}}, \bibinfo{pages}{2204--2212}
  (\bibinfo{publisher}{Curran Associates, Inc.}, \bibinfo{year}{2014}).

\bibitem{Karpathy2015-it}
\bibinfo{author}{Karpathy, A.} \& \bibinfo{author}{Fei-Fei, L.}
\newblock \bibinfo{title}{Deep visual-semantic alignments for generating image
  descriptions}.
\newblock In \emph{\bibinfo{booktitle}{Proceedings of the {IEEE} conference on
  computer vision and pattern recognition}}, \bibinfo{pages}{3128--3137}
  (\bibinfo{publisher}{cv-foundation.org}, \bibinfo{year}{2015}).

\bibitem{Zhang2016-yq}
\bibinfo{author}{Zhang, R.}, \bibinfo{author}{Isola, P.} \&
  \bibinfo{author}{Efros, A.~A.}
\newblock \bibinfo{title}{Colorful image colorization}.
\newblock In \emph{\bibinfo{booktitle}{Computer Vision -- {ECCV} 2016}},
  \bibinfo{pages}{649--666} (\bibinfo{publisher}{Springer International
  Publishing}, \bibinfo{year}{2016}).

\bibitem{Dong2014-iy}
\bibinfo{author}{Dong, C.}, \bibinfo{author}{Loy, C.~C.}, \bibinfo{author}{He,
  K.} \& \bibinfo{author}{Tang, X.}
\newblock \bibinfo{title}{Learning a deep convolutional network for image
  {Super-Resolution}}.
\newblock In \emph{\bibinfo{booktitle}{Computer Vision -- {ECCV} 2014}},
  \bibinfo{pages}{184--199} (\bibinfo{publisher}{Springer International
  Publishing}, \bibinfo{year}{2014}).

\bibitem{Wang2015-va}
\bibinfo{author}{Wang, L.}, \bibinfo{author}{Ouyang, W.},
  \bibinfo{author}{Wang, X.} \& \bibinfo{author}{Lu, H.}
\newblock \bibinfo{title}{Visual tracking with fully convolutional networks}.
\newblock In \emph{\bibinfo{booktitle}{Proceedings of the {IEEE} international
  conference on computer vision}}, \bibinfo{pages}{3119--3127}
  (\bibinfo{publisher}{cv-foundation.org}, \bibinfo{year}{2015}).

\bibitem{Girshick2014-cv}
\bibinfo{author}{Girshick, R.}, \bibinfo{author}{Donahue, J.},
  \bibinfo{author}{Darrell, T.} \& \bibinfo{author}{Malik, J.}
\newblock \bibinfo{title}{Rich feature hierarchies for accurate object
  detection and semantic segmentation}.
\newblock In \emph{\bibinfo{booktitle}{Proceedings of the {IEEE} conference on
  computer vision and pattern recognition}}, \bibinfo{pages}{580--587}
  (\bibinfo{publisher}{cv-foundation.org}, \bibinfo{year}{2014}).

\bibitem{Kraus2018-su}
\bibinfo{author}{Krau{\ss}, S.~D.} \emph{et~al.}
\newblock \bibinfo{title}{Hierarchical deep convolutional neural networks
  combine spectral and spatial information for highly accurate
  raman‐microscopy‐based cytopathology}.
\newblock \emph{\bibinfo{journal}{J. Biophotonics}}
  \textbf{\bibinfo{volume}{11}}, \bibinfo{pages}{e201800022}
  (\bibinfo{year}{2018}).

\bibitem{Lotfollahi2019-bf}
\bibinfo{author}{Lotfollahi, M.}, \bibinfo{author}{Berisha, S.},
  \bibinfo{author}{Daeinejad, D.} \& \bibinfo{author}{Mayerich, D.}
\newblock \bibinfo{title}{Digital staining of {High-Definition} fourier
  transform infrared ({FT-IR}) images using deep learning}.
\newblock \emph{\bibinfo{journal}{Appl. Spectrosc.}}
  \textbf{\bibinfo{volume}{73}}, \bibinfo{pages}{556--564}
  (\bibinfo{year}{2019}).

\bibitem{Berisha2019-gr}
\bibinfo{author}{Berisha, S.} \emph{et~al.}
\newblock \bibinfo{title}{Deep learning for {FTIR} histology: leveraging
  spatial and spectral features with convolutional neural networks}.
\newblock \emph{\bibinfo{journal}{Analyst}} \textbf{\bibinfo{volume}{144}},
  \bibinfo{pages}{1642--1653} (\bibinfo{year}{2019}).

\bibitem{Kampe2017-ue}
\bibinfo{author}{Kampe, B.}, \bibinfo{author}{Klo{\ss}, S.},
  \bibinfo{author}{Bocklitz, T.}, \bibinfo{author}{R{\"o}sch, P.} \&
  \bibinfo{author}{Popp, J.}
\newblock \bibinfo{title}{Recursive feature elimination in raman spectra with
  support vector machines}.
\newblock \emph{\bibinfo{journal}{Front. Optoelectron.}}
  \textbf{\bibinfo{volume}{10}}, \bibinfo{pages}{273--279}
  (\bibinfo{year}{2017}).

\bibitem{Guo2018-iw}
\bibinfo{author}{Guo, S.} \emph{et~al.}
\newblock \bibinfo{title}{Model transfer for raman-spectroscopy-based bacterial
  classification}.
\newblock \emph{\bibinfo{journal}{J. Raman Spectrosc.}}
  \textbf{\bibinfo{volume}{49}}, \bibinfo{pages}{627--637}
  (\bibinfo{year}{2018}).

\bibitem{Gurbani2018-da}
\bibinfo{author}{Gurbani, S.~S.} \emph{et~al.}
\newblock \bibinfo{title}{A convolutional neural network to filter artifacts in
  spectroscopic {MRI}}.
\newblock \emph{\bibinfo{journal}{Magn. Reson. Med.}}  (\bibinfo{year}{2018}).

\bibitem{Malek2018-cp}
\bibinfo{author}{Malek, S.}, \bibinfo{author}{Melgani, F.} \&
  \bibinfo{author}{Bazi, Y.}
\newblock \bibinfo{title}{One-dimensional convolutional neural networks for
  spectroscopic signal regression: Feature extraction based on {1D-CNN} is
  proposed and validated}.
\newblock \emph{\bibinfo{journal}{J. Chemom.}} \textbf{\bibinfo{volume}{32}},
  \bibinfo{pages}{e2977} (\bibinfo{year}{2018}).

\bibitem{Liu2017-wk}
\bibinfo{author}{Liu, J.} \emph{et~al.}
\newblock \bibinfo{title}{Deep convolutional neural networks for raman spectrum
  recognition: A unified solution}.
\newblock \emph{\bibinfo{journal}{Analyst}}  (\bibinfo{year}{2017}).

\bibitem{Zhang2019-ju}
\bibinfo{author}{Zhang, X.}, \bibinfo{author}{Lin, T.}, \bibinfo{author}{Xu,
  J.}, \bibinfo{author}{Luo, X.} \& \bibinfo{author}{Ying, Y.}
\newblock \bibinfo{title}{{DeepSpectra}: An end-to-end deep learning approach
  for quantitative spectral analysis}.
\newblock \emph{\bibinfo{journal}{Anal. Chim. Acta}}
  \textbf{\bibinfo{volume}{1058}}, \bibinfo{pages}{48--57}
  (\bibinfo{year}{2019}).

\bibitem{He2015-zl}
\bibinfo{author}{He, K.}, \bibinfo{author}{Zhang, X.}, \bibinfo{author}{Ren,
  S.} \& \bibinfo{author}{Sun, J.}
\newblock \bibinfo{title}{Deep residual learning for image recognition}.
\newblock In \emph{\bibinfo{booktitle}{Proceedings of the IEEE conference on
  computer vision and pattern recognition}}, \bibinfo{pages}{770--778}
  (\bibinfo{year}{2016}).

\bibitem{Dumoulin2016-bx}
\bibinfo{author}{{Dumoulin}, V.} \& \bibinfo{author}{{Visin}, F.}
\newblock \bibinfo{title}{{A guide to convolution arithmetic for deep learning.
  Preprint at https://arxiv.org/abs/1603.07285 (2016)}} .

\bibitem{Banaei_undated-ux}
\bibinfo{author}{Banaei, N.}, \bibinfo{author}{Watz, N.},
  \bibinfo{author}{Getsinger, D.} \& \bibinfo{author}{Ghafghaichi, L.}
\newblock \bibinfo{title}{{SUH} antibiogram data for bacterial and yeast
  isolates}.
\newblock \bibinfo{type}{Tech. Rep.}, \bibinfo{institution}{Stanford Healthcare
  Clinical Microbiology Laboratory} (\bibinfo{year}{2016}).
\newblock
  \urlprefix\url{http://med.stanford.edu/bugsanddrugs/clinical-microbiology/_jcr_content/main/panel_builder/panel_0/download_748639600/file.res/SHC\%20antibiogram\202016.pdf}.

\bibitem{Lamy2016-vs}
\bibinfo{author}{Lamy, B.}, \bibinfo{author}{Darg{\`e}re, S.},
  \bibinfo{author}{Arendrup, M.~C.}, \bibinfo{author}{Parienti, J.-J.} \&
  \bibinfo{author}{Tattevin, P.}
\newblock \bibinfo{title}{How to optimize the use of blood cultures for the
  diagnosis of bloodstream infections? a state-of-the art}.
\newblock \emph{\bibinfo{journal}{Front. Microbiol.}}
  \textbf{\bibinfo{volume}{7}}, \bibinfo{pages}{697} (\bibinfo{year}{2016}).

\bibitem{Reimer1997-rr}
\bibinfo{author}{Reimer, L.~G.}, \bibinfo{author}{Wilson, M.~L.} \&
  \bibinfo{author}{Weinstein, M.~P.}
\newblock \bibinfo{title}{Update on detection of bacteremia and fungemia}.
\newblock \emph{\bibinfo{journal}{Clin. Microbiol. Rev.}}
  \textbf{\bibinfo{volume}{10}}, \bibinfo{pages}{444--465}
  (\bibinfo{year}{1997}).

\bibitem{Kogler2018-pf}
\bibinfo{author}{K{\"o}gler, M.} \emph{et~al.}
\newblock \bibinfo{title}{Bare laser-synthesized au-based nanoparticles as
  nondisturbing surface-enhanced raman scattering probes for bacteria
  identification}.
\newblock \emph{\bibinfo{journal}{J. Biophotonics}}
  \textbf{\bibinfo{volume}{11}}, \bibinfo{pages}{e201700225}
  (\bibinfo{year}{2018}).

\bibitem{Chen2018-zr}
\bibinfo{author}{Chen, Y.}, \bibinfo{author}{Premasiri, W.~R.} \&
  \bibinfo{author}{Ziegler, L.~D.}
\newblock \bibinfo{title}{Surface enhanced raman spectroscopy of chlamydia
  trachomatis and neisseria gonorrhoeae for diagnostics, and extra-cellular
  metabolomics and biochemical monitoring}.
\newblock \emph{\bibinfo{journal}{Sci. Rep.}} \textbf{\bibinfo{volume}{8}},
  \bibinfo{pages}{5163} (\bibinfo{year}{2018}).

\bibitem{Li2010-oc}
\bibinfo{author}{Li, J.~F.} \emph{et~al.}
\newblock \bibinfo{title}{Shell-isolated nanoparticle-enhanced raman
  spectroscopy}.
\newblock \emph{\bibinfo{journal}{Nature}} \textbf{\bibinfo{volume}{464}},
  \bibinfo{pages}{392--395} (\bibinfo{year}{2010}).

\bibitem{Cronquist2012-uu}
\bibinfo{author}{Cronquist, A.~B.} \emph{et~al.}
\newblock \bibinfo{title}{Impacts of culture-independent diagnostic practices
  on public health surveillance for bacterial enteric pathogens}.
\newblock \emph{\bibinfo{journal}{Clin. Infect. Dis.}}
  \textbf{\bibinfo{volume}{54 Suppl 5}}, \bibinfo{pages}{S432--9}
  (\bibinfo{year}{2012}).

\bibitem{Kang2014-vy}
\bibinfo{author}{Kang, D.-K.} \emph{et~al.}
\newblock \bibinfo{title}{Rapid detection of single bacteria in unprocessed
  blood using integrated comprehensive droplet digital detection}.
\newblock \emph{\bibinfo{journal}{Nat. Commun.}} \textbf{\bibinfo{volume}{5}},
  \bibinfo{pages}{5427} (\bibinfo{year}{2014}).

\bibitem{Tung2017-uo}
\bibinfo{author}{Tung, P.-Y.} \emph{et~al.}
\newblock \bibinfo{title}{Batch effects and the effective design of single-cell
  gene expression studies}.
\newblock \emph{\bibinfo{journal}{Sci. Rep.}} \textbf{\bibinfo{volume}{7}},
  \bibinfo{pages}{39921} (\bibinfo{year}{2017}).

\bibitem{Wang2015-fd}
\bibinfo{author}{Wang, Y.} \& \bibinfo{author}{Navin, N.~E.}
\newblock \bibinfo{title}{Advances and applications of single-cell sequencing
  technologies}.
\newblock \emph{\bibinfo{journal}{Mol. Cell}} \textbf{\bibinfo{volume}{58}},
  \bibinfo{pages}{598--609} (\bibinfo{year}{2015}).

\bibitem{Pallen2010-jo}
\bibinfo{author}{Pallen, M.~J.}, \bibinfo{author}{Loman, N.~J.} \&
  \bibinfo{author}{Penn, C.~W.}
\newblock \bibinfo{title}{High-throughput sequencing and clinical microbiology:
  progress, opportunities and challenges}.
\newblock \emph{\bibinfo{journal}{Curr. Opin. Microbiol.}}
  \textbf{\bibinfo{volume}{13}}, \bibinfo{pages}{625--631}
  (\bibinfo{year}{2010}).

\bibitem{Chung2015-zk}
\bibinfo{author}{Chung, J.}, \bibinfo{author}{Kang, J.~S.},
  \bibinfo{author}{Jurng, J.~S.}, \bibinfo{author}{Jung, J.~H.} \&
  \bibinfo{author}{Kim, B.~C.}
\newblock \bibinfo{title}{Fast and continuous microorganism detection using
  aptamer-conjugated fluorescent nanoparticles on an optofluidic platform}.
\newblock \emph{\bibinfo{journal}{Biosens. Bioelectron.}}
  \textbf{\bibinfo{volume}{67}}, \bibinfo{pages}{303--308}
  (\bibinfo{year}{2015}).

\bibitem{Diep2008-op}
\bibinfo{author}{Diep, B.~A.} \emph{et~al.}
\newblock \bibinfo{title}{The arginine catabolic mobile element and
  staphylococcal chromosomal cassette mec linkage: convergence of virulence and
  resistance in the {USA300} clone of methicillin-resistant staphylococcus
  aureus}.
\newblock \emph{\bibinfo{journal}{J. Infect. Dis.}}
  \textbf{\bibinfo{volume}{197}}, \bibinfo{pages}{1523--1530}
  (\bibinfo{year}{2008}).

\bibitem{Kingma2014-ni}
\bibinfo{author}{Kingma, D.~P.} \& \bibinfo{author}{Ba, J.}
\newblock \bibinfo{title}{Adam: {A} method for stochastic optimization.
  {P}reprint at https://arxiv.org/abs/1412.6980. (2014)} .

\bibitem{Nakasone2017-eb}
\bibinfo{author}{Nakasone, T.} \emph{et~al.}
\newblock \bibinfo{title}{Bacterial susceptibility report: 2016}.
\newblock \bibinfo{type}{Tech. Rep.}, \bibinfo{institution}{VA Palo Alto Health
  Care System} (\bibinfo{year}{2017}).
\newblock
  \urlprefix\url{https://web.stanford.edu/~jonc101/tools/Antibiogram/VAPAabgm2016Report\%20FINAL4-14-17.pdf}.

\bibitem{Lorenz2017-lt}
\bibinfo{author}{Lorenz, B.}, \bibinfo{author}{Wichmann, C.},
  \bibinfo{author}{St{\"o}ckel, S.}, \bibinfo{author}{R{\"o}sch, P.} \&
  \bibinfo{author}{Popp, J.}
\newblock \bibinfo{title}{{Cultivation-Free} raman spectroscopic investigations
  of bacteria}.
\newblock \emph{\bibinfo{journal}{Trends Microbiol.}}
  \textbf{\bibinfo{volume}{25}}, \bibinfo{pages}{413--424}
  (\bibinfo{year}{2017}).

\end{thebibliography}
\baselineskip24pt

\newpage
\clearpage
\pagebreak

\begin{addendum}
 \item [Acknowledgements]
The authors gratefully acknowledge the assistance of Joel Jean, Chi-Min Ho, Alice Lay, Katherine Sytwu, Randy Mehlenbacher, Tracey Hong, Samuel Lee, David Zeng, Mark Winters, Marcin Walkiewicz and Andrey Malkovskiy.
Raman measurements were performed at the Stanford Nano Shared Facilities (SNSF), supported by the National Science Foundation under award ECCS-1542152.
The authors gratefully acknowledge support from the Alfred P. Sloan Foundation, the Stanford Catalyst for Collaborative Solutions and the Gates Foundation. N.J. acknowledges support from the Department of Defense (DoD) through the National Defense Science \& Engineering Graduate Fellowship (NDSEG) Program.

 \item [Author contributions]
C.H. and N.J. conceptualized the algorithms, analyzed classification results, and fine-tuned the algorithms. C.H. developed sample preparation and data collection protocols, and collected the datasets. N.J. designed, optimized, and trained the algorithms.
C.H. and L.B. prepared sample cultures.
N.B., M.H., and C.A.H. developed the antibiotic groupings, collected samples, and provided input on clinical relevance.
A.A.E.S, N.B., and J.A.D. conceived the initial idea and C.H. and N.J. further developed the idea.
J.A.D. and A.A.E.S. supervised the project along with supervision from S.S.J., N.B., M.H., and S.E. on relevant portions of the research.
All authors contributed to editing of the manuscript.

 \item[Competing Interests] The authors declare no competing interests.

\end{addendum}

\end{document}